\documentclass[preprint,aps,amsmath,superscriptaddress,nofootinbib,tightenlines]{revtex4}
\usepackage{graphicx}
\usepackage{bm}
\usepackage{epsfig}

\newif\ifpdf
\ifx\pdfoutput\undefined
\pdffalse 
\else
\pdfoutput=1 
\pdftrue
\fi

\def\bsigma{\mbox{\boldmath $\sigma$}}
\def\bomega{\mbox{\boldmath $\omega$}}

\def\OMIT#1{}

\newcommand{\nn}{\nonumber} 

\newcommand{\bn}{{\bar n}}
\newcommand{\bea}{\begin{eqnarray}}
\newcommand{\eea}{\end{eqnarray}}

\newcommand{\bnP}{\bar {\cal P}}
\newcommand{\ppP}{{\cal P}_\perp}

\newcommand{\cP}{{\cal P}}

\newcommand{\LQCD}{{\Lambda_{\rm QCD}}}

\newcommand{\jpsi}{J/\psi}
\newcommand{\eps}{\epsilon}

\newcommand{\SCETa}{SCET$_{\rm I} \,\,$}
\newcommand{\SCETaa}{SCET$_{\rm I}$}
\newcommand{\SCETb}{SCET$_{\rm II} \,\,$}
\newcommand{\SCETbb}{SCET$_{\rm II}$}

\begin{document}

\ifpdf
\DeclareGraphicsExtensions{.pdf, .jpg}
\else
\DeclareGraphicsExtensions{.eps, .jpg,.ps}
\fi


\title{Resummation of Large Endpoint Corrections to Color-Octet $J/\psi$ Photoproduction} 

\author{Sean Fleming\footnote{Electronic address: fleming@physics.arizona.edu}}
\affiliation{Physics Department, University of Arizona,
      	Tucson, AZ 85721
	\vspace{0.1cm}}
	
\author{Adam K. Leibovich\footnote{Electronic address: akl2@pitt.edu}}
\affiliation{Department of Physics and Astronomy, 
	University of Pittsburgh,
        Pittsburgh, PA 15260
\vspace{0.2cm}}

\author{Thomas Mehen\footnote{Electronic address: mehen@phy.duke.edu}}
\affiliation{Department of Physics, Duke University, Durham,  NC 27708\vspace{0.2cm}}
\affiliation{Jefferson Laboratory, 12000 Jefferson Ave., Newport News, VA 23606\vspace{0.2cm}}
\date{\today\\ \vspace{1cm} }



\begin{abstract} 

An unresolved problem in $J/\psi$ phenomenology is a systematic understanding of the
differential photoproduction cross section, $d\sigma/dz [\gamma + p \to J/\psi + X$], where $z=
E_\psi/E_\gamma$ in the proton rest frame. In the non-relativistic QCD (NRQCD) factorization
formalism,  fixed-order perturbative calculations of color-octet mechanisms   suffer from large
perturbative and nonperturbative corrections that grow rapidly in the endpoint region, $z \to 1$. In this
paper, NRQCD and soft collinear effective theory  are combined to resum these large
 corrections to the color-octet photoproduction cross section.
We derive a factorization theorem for the endpoint differential cross section involving the
parton distribution function and the color-octet $\jpsi$ shape functions. A one loop
matching calculation  explicitly confirms our factorization theorem at next-to-leading order. 
Large perturbative corrections are resummed using the  renormalization group. The calculation of
the color-octet contribution to  $d\sigma/dz$ is in qualitative  agreement with data.
Quantitative tests of the universality of color-octet matrix  elements require improved
knowledge of shape functions entering these calculations as well as resummation of the
color-singlet contribution which accounts for much of the total cross section and  also peaks
near the endpoint. 

\end{abstract}

\maketitle

\newpage
\section{Introduction}

Our current understanding of the production of heavy quarkonia is based on Non-Relativistic Quantum Chromodynamics
(NRQCD)~\cite{Caswell:1985ui,bbl, Luke:2000kz}, an effective theory for bound states of two or more heavy quarks. The NRQCD
factorization formalism solves important theoretical and phenomenological problems in quarkonium production and decay. Early
color-singlet model calculations of $\chi_c$ decay were plagued by infrared divergences \cite{Barbieri:1976fp}. NRQCD solves  this
problem by  providing a generalized factorization theorem which allows for infrared safe calculations of inclusive production and
decay rates~\cite{Bodwin:1992ye}. This formalism incorporates nonperturbative corrections to the color-singlet model, including
color-octet decay and production mechanisms. Color-octet production mechanisms are necessary for understanding the production of
$J/\psi$ at large transverse momentum at the Fermilab Tevatron \cite{tevatron,choleibovich}. However, the polarization of the observed
$J/\psi$ remains poorly understood~\cite{Affolder:2000nn}.  

The color-octet contribution to  $J/\psi$ photoproduction gives a large enhancement to the cross section near the kinematic endpoint
defined by $z \to 1$, where $z=E_{\psi}/E_\gamma$ in the proton rest frame~\cite{Cacciari:1996dg,Ko:1996xw}. Ref.~\cite{Cacciari:1996dg} proposed that
this peak constitutes a distinct signal for a color-octet contribution to photoproduction, and observed that such a signal is in contradiction
with the experimental data which do not exhibit a peak. A crucial test of the NRQCD factorization theorem is verifying the universality of the color-octet matrix elements  which enter into
a variety of different $J/\psi$ production processes, and the unobserved excess of $J/\psi$ at the photoproduction endpoint
could be interpreted as a failure of universality.

Color-octet matrix elements were first fit to hadroproduction data from the Tevatron~\cite{choleibovich}, and have been refitted
a number of times~\cite{Cacciari:1995yt,Beneke:1996yw,Kniehl:1998qy,Braaten:1999qk} with  general agreement among the various results. In
general the extraction of the color-octet matrix elements has fairly large theoretical errors associated with it. The errors are
particularly large on the extracted values of the color-octet matrix elements $\langle{\cal O}_8^{\jpsi}(^1S_0)\rangle$ and 
$\langle{\cal O}_8^{\jpsi}(^3P_J)\rangle$, which dominate the color-octet contribution to photoproduction, where a variation  of almost an order of magnitude is typical. These large errors are thought
to be due to higher order QCD corrections and as a result there has been an attempt to incorporate a subset of higher order contributions
through the use of Monte Carlo methods~\cite{Cano-Coloma:1997rn} and a $k_T$ factorization approach~\cite{Baranov:2002cf}. These
approaches extract values for color-octet matrix elements up to 5 times smaller than the fixed-order calculations. Unfortunately, 
the rise in the fixed-order calculation of the color-octet photoproduction differential cross section is so dramatic that even the large 
uncertainties in the color-octet matrix elements cannot account for the discrepancy.

However, as one approaches the endpoint region, $z \to 1$, the fixed-order calculation 
of Refs.~\cite{Cacciari:1996dg,Ko:1996xw} is invalidated by large perturbative and nonperturbative
corrections. Specifically, perturbative corrections of the form
\begin{equation}
\alpha_s^n \frac{ {\rm ln}^m(1-z)}{1-z}   \quad , \quad m \leq 2 n - 1\nonumber
\end{equation}
are the source of the growth of the cross section as $z \to 1$~\cite{Kniehl:1997gh,Kniehl:1997fv}. These corrections  
must be resummed to all orders to make sensible comparison with data. There
are also nonperturbative corrections scaling like $v^{2n}/(1-z)^n$, where $v$ is the typical 
velocity of the $c \bar c$ within the $J/\psi$~\cite{Beneke:1997qw}. These corrections 
invalidate the NRQCD expansion near the kinematic endpoint. Ref.~\cite{Beneke:1999gq} addressed
the latter issue by recalculating $J/\psi$ photoproduction including a nonperturbative 
shape function which resums these corrections. The shape function tames perturbative
endpoint divergences and results in a differential cross section that peaks near $z \approx 0.9$.
However, the peak in the differential cross section is too narrow to be compatible with data,
and the authors conclude that a resummation of the singular perturbative contributions
must also be carried out. 

Similar issues arise when analyzing the production of $J/\psi$ in $e^+e^-$ collisions. In Ref.\cite{Fleming:2003gt} a
color-octet factorization theorem for $e^+ + e^- \to J/\psi + X$ was derived using a combination of NRQCD and soft collinear
effective theory (SCET)~\cite{Bauer:2001ew,Bauer:2001yr,Bauer:2001ct,Bauer:2001yt}. This approach resums nonperturbative
and perturbative corrections that are enhanced near the kinematic endpoint.   An important conclusion of
Ref.~\cite{Fleming:2003gt} is that both the resummed perturbative corrections and the resummed non-perturbative corrections
are necessary because the two effects act  constructively to significantly broaden the spectrum.   

In this paper we apply NRQCD and SCET to photoproduction at the endpoint of the spectrum.  We derive a new factorization
theorem for color-octet  $J/\psi$ photoproduction which resums endpoint corrections for this process. The factorization
theorem in this case is more complicated than that for $e^+ + e^- \to J/\psi + X$ because there is a parton in the initial
state hadron. To derive the factorization theorem, we first match QCD onto \SCETa (in which collinear partons  have
off-shellness $O(\sqrt{M\Lambda_{\rm QCD}})$) at the scale $M = 2 m_c$, 
and then match \SCETa onto \SCETb (where collinear partons have
off-shellness $O(\Lambda_{\rm QCD})$)~\cite{Bauer:2002aj}  at an intermediate scale, $M \sqrt{1-z}$. At each stage we check that the effective
field theory correctly reproduces the infrared physics of the previous theory to ensure that large logarithms are correctly
resummed. Similarly, evolution of the renormalization group equations (RGE) is carried out in two stages. The
first stage of evolution, from the scale $M$ to the intermediate scale, $M\sqrt{1-z}$, is performed in \SCETa,
while the second stage of evolution, from $M \sqrt{1-z}$ to $M (1-z)$, involves \SCETb running.

The factorization theorem is novel in that the $J/\psi$ are not required to be produced with large transverse momentum, $p_\perp$,
with respect to the photon-proton beam axis. Most perturbative analyses of $J/\psi$ photoproduction assume that
$p_\perp$ of $O(1 \,{\rm GeV})$ or larger is required for pQCD to be applicable to this process. For this reason, most
phenomenological and experimental analyses impose a $p_\perp$ cut on the data. Such a cut is inapproriate for application
of the factorization theorem and the resummed cross sections derived in this paper. 

Our final result for the color-octet contribution to $d\sigma/dz$ exhibits a spectrum that is significantly broadened 
and has a peak reduced in height relative to including only  a shape function or only resumming the perturbative
corrections, as anticipated. The inclusion of both nonperturbative and perturbative corrections significantly broadens
the spectrum and is in qualitative agreement with data. However, fixed order color-singlet  calculations are
consistent with existing data on photoproduction so if these calculations are naively combined with our calculation,
the color-octet matrix elements in photoproduction need to be roughly an order of magnitude smaller than those
extracted from fixed-order analyses of Tevatron data.  However, there are several sources of uncertainty which make
it premature to attempt quantitative extraction of color-octet matrix elements using our calculation.
Currently, the shape functions appearing in the photoproduction calculation have not been precisely determined. Also,
consistency requires a similar resummation of endpoint effects in  color-singlet photoproduction which accounts for
much of the total photoproduction cross section. Finally, all of the existing data has cuts on $p_\perp$
and/or cuts on diffractive contributions which are inappropriate for our analysis. Since the resummation 
is expected to suppress the color-singlet contribution, and the cuts tends to reduce the experimental cross section
at largest values of $z$, a future analysis which takes these two effects into account could find more room for a
color-octet contribution.

This paper is organized as follows. In Section II, we derive the factorization theorem for the color-octet $\jpsi$ photoproduction cross
section near the endpoint, using the two-stage matching procedure discussed earlier. In Section III, we obtain analytic expressions for the
leading logarithmically enhanced corrections at next-to-leading order (NLO) in QCD.  In Section IV, we perform the matching calculations.
This section is broken up into two subsections. In the first subsection we match the \SCETa current onto the QCD amplitude for $\gamma + g
\to c \bar{c}$ at one loop and show that large logarithms are minimized at the scale $M$. In the second subsection we first calculate the
NLO cross section in \SCETa.  Correct evaluation of the \SCETa cross section requires care in avoiding the double counting of usoft modes 
and collinear modes with vanishing label momentum, the so-called ``zero-bin'' modes, as recently discussed in Ref.~\cite{Manohar:2006nz}.
Taking this subtlety into account is necessary to demonstrate that \SCETa  reproduces the large endpoint corrections of the NLO QCD
calculation  extracted in section III. We then match the \SCETa result onto \SCETbb, which determines the intermediate matching scale. In
Section V, we perform the RGE evolution which resums large perturbative corrections.  In Section VI, we discuss the impact of the resummed
color-octet differential cross section on the phenomenology of $\jpsi$ photoproduction. A brief summary is provided in Section VII.
Appendix A derives formulae that are useful for the
extracting the large NLO corrections in Section III, and Appendix B provides some details about the zero-bin subtractions. Preliminary results describing some of the work here were presented at the Ringberg Workshop on New Trends in HERA Physics 2005~\cite{Fleming:2005pt}.

\section{Factorization}

In this section we derive a factorization theorem valid for $J/\psi$ photoproduction near the kinematic  endpoint, which is defined
by $z\sim 1$, where $z = p_\psi\!\cdot\!p_P / p_\gamma\!\cdot\!p_P$, and $p_\psi$, $p_P$ and $p_\gamma$ are the four-momenta of the 
$\jpsi$, the initial state proton, and initial state photon, respectively. In the proton rest frame  $z = E_\psi/ E_\gamma$. We begin
by showing that in the limit  $z \to 1$, NRQCD factorization breaks down and SCET is required to obtain the appropriate factorization
formula.

In the proton-photon center-of-mass frame
\begin{equation}
p_\gamma^\mu = \frac{\sqrt{s}}{2} \bar n^\mu, \qquad 
p_P^\mu = \frac{\sqrt{s}}{2} n^\mu, \qquad
p_{c\bar{c}}^\mu = M v^\mu + k^\mu \,,
\end{equation}
where $s=(p_\gamma + p_P)^2$, $n^\mu = (1,0,0,-1)$, $\bn^\mu = (1,0,0,1)$, $M = 2m_c$, and $v^\mu$ and $k^\mu$ are the 4-velocity of
the $J/\psi$ and the residual momentum of the $c\bar c$ pair in the $J/\psi$, respectively.  In terms of the scaling variable, $z$,
the $\jpsi$ velocity is
\begin{equation}
p_\psi^\mu = M_\psi v^\mu = \frac{z\sqrt{s}}{2} \bar n^\mu + p_\perp^\mu + \frac{m_\perp^2}{2 z \sqrt{s}}n^\mu \, ,
\end{equation}
where $m_\perp^2 = M_\psi^2 + {\bf p}_\perp^2$.
By momentum conservation we have
\bea
p_X^\mu &=& p_\gamma^\mu + p_P^\mu - p_{c\bar{c}}^\mu 
\nn \\
&=& \frac{\sqrt{s}}{2}  \bigg( 1 - \frac{M}{M_\psi} z \bigg) \bn^\mu 
+ \frac{\sqrt{s}}{2}  \bigg( 1 -  \frac{Mm^2_\perp}{sz M_\psi } \bigg) n^\mu
- \frac{M}{M_\psi}p^\mu_\perp-k^\mu  \,.
\eea
In NRQCD $k^\mu$ is typically dropped at lowest order, and terms with powers of $k^\mu$ are matched onto NRQCD operators with derivatives that are higher
order  in the $v$ expansion. When the kinematics of a quarkonium production or decay process become  sensitive to any components of
$k^\mu$ such that this expansion breaks down, resummation of NRQCD operators into a shape function is required
\cite{Beneke:1997qw}.

In the rest frame of the $J/\psi$ all components of $k^\mu$ scale as $\Lambda_{\rm QCD}$. 
In the proton-photon center-of-mass frame, the components
of $k^\mu$ scale as 
\begin{equation}
k^\mu = (n\cdot k, \bn \cdot k ,k_\perp) \sim 
\Lambda_{\rm QCD} \left( \frac{z \sqrt{s}}{M}, \frac{M}{z\sqrt{s}}, 1\right) \, ,
\end{equation}
where we have neglected $O(p_\perp/\sqrt{s})$ corrections which are small in the endpoint region.
The $n \cdot k$ component of $k^\mu$ is enhanced by $\sqrt{s}/M$, while the $\bn \cdot k$  
is suppressed by the same amount. Computing $p_X^2$ 
and keeping only the  $n \cdot k$ component of $k^\mu$ we find
\begin{equation}\label{invmass}
p^2_X =  s \bigg( 1 - \frac{M}{M_\psi} z \bigg)\bigg( 1 -  \frac{Mm^2_\perp}{sz M_\psi } \bigg) 
- \frac{M^2}{M_\psi^2} {\bf p}_\perp^2- \sqrt{s}  \bigg( 1 -  \frac{Mm^2_\perp}{sz M_\psi } \bigg) n \cdot k 
+ ... \, .
\end{equation}
In the endpoint region, the transverse momentum of the $J/\psi$ pair is of order $\sqrt{\hat{s}(1-z)} \sim 
\sqrt{\LQCD M} \sim 1 \, \textrm{GeV}$, where $\sqrt{\hat{s}}$ is the {\em partonic} center-of-mass energy which is of order $M$
in the endpoint region. Therefore, the term in Eq.~(\ref{invmass}) proportional to ${\bf p}_\perp^2$ is unimportant. Furthermore, 
for HERA kinematics, $\sqrt{s} \sim 100$ GeV, so $M m^2_\perp/(s z M_\psi) \sim M^2/s \sim 10^{-3}$ for $z \approx 1$. 
The last term in Eq.~(\ref{invmass}) is as important as the first term when $n\cdot k  \sim \sqrt{s}(1-M z/M_\psi)$,
which occurs when $z\sim (M_\psi -\LQCD)/M \sim 1$.  In this region NRQCD factorization breaks down.
>From the point of view of SCET, a new factorization theorem is required when the 
final state particles recoiling against the $J/\psi$ are jet-like and have to  be described by SCET collinear fields rather than   integrated out as in conventional NRQCD factorization. The final state is jet-like when $\bn \cdot p_X \gg \sqrt{p_X^2}$, which is equivalent to $(1-M z/M_\psi)  \ll 1$. This  leads
to the conclusion that as $z \to 1$,  SCET is required for  $J/\psi$ photoproduction. 

We derive the factorization formula for the photoproduction cross section in the endpoint region in two steps. First we match the QCD
amplitude for $\gamma + g \to c\bar{c}$ onto the SCET${}_{\rm I}$ current:
\begin{equation}\label{currmatch}
J^\mu(x) =\int  d^3{\bomega} \,
e^{i(Mv - \bomega )\cdot x} C^{\mu}_{ \alpha}(\bomega) 
J^\alpha(\bomega,  x) \,,
\end{equation}
where 
\bea
\int d^3{\bomega}  \equiv \int d \bar{\omega} \, d^2 \omega_\perp,
\eea
with $\bomega^\mu = \bar{\omega} n^\mu/2 + \omega^\mu_\perp$. The leading order contribution  from the 
color-octet  $^1S_0$ current is 
\begin{equation}
J^\alpha (\bomega, x)  = 
\big[ \psi^\dagger_{{\bf p}} \delta^{(3)}({\vec{\cP} - \bomega})B^\alpha_{\perp} \chi_{-{\bf p}} \big] (x) \,,
\end{equation}
%
where $B_\perp^\alpha$ is the gauge invariant collinear gluon field defined by 
\bea
B_\perp^\alpha = \frac{1}{g_s} W^\dagger({\cal P}_\perp^\alpha + g_s (A^\alpha_{n,q})_\perp) W \, ,\nn
\eea
and $\psi_{{\bf p}}$ and $\chi_{{\bf p}}$
are the NRQCD fields for the heavy quarks and antiquarks, respectively. The
color-octet $^3P_J$ current is
\begin{equation}
J^\alpha_{\sigma \delta} (\bomega , x)  = 
\Lambda\cdot\widehat{\bf p}_\sigma  
[\psi^\dagger_{{\bf p}} \Lambda\cdot\bsigma_\delta \delta^{(3)}({\vec{\cP} - \bomega})B^\alpha_{\perp}  
\chi_{-{\bf p}}] \,,
\end{equation}
where $\widehat{\bf p}_\sigma = \Lambda_{i \sigma} p^i / M$, $\Lambda^{\mu\nu}$ is the boost matrix from the lab frame to the 
$c\bar{c}$ rest frame, and $\vec{\cP}^\mu = \bnP n^\mu/2 + \ppP^\mu$ is the operator which projects out label momentum. We have
defined
\begin{equation}
\delta^{(3)}({\vec{\cP} - \bomega}) \equiv \delta(\bnP - \bar{\omega}) \delta^{(2)}(\ppP-\omega_\perp) \,.
\end{equation}
Tree level matching of QCD onto SCET${}_{\rm I}$ determines the leading order 
Wilson coefficients for the $^1S_0^{(8)}$ and $^3P_J^{(8)}$ channels
~\cite{Fleming:2003gt}:
\bea\label{match}
C^{\mu \alpha} (^1S_0^{(8)})&=& \frac{- 2 e e_c g_s(M)}{M} \epsilon_\perp^{\mu \alpha} \,,
\nonumber \\
C^{\mu \alpha\delta\sigma} (^3P_J^{(8)})&=& \frac{- i 4 e e_c g_s(M)}{M} \left(g_\perp^{\alpha\delta} g_\perp^{\mu\sigma} +
g_\perp^{\alpha\sigma} g_\perp^{\mu\delta} - g_\perp^{\alpha\mu} \bar{n}^\sigma \bar{n}^\delta 
 \right) \, .
\eea   

The second step is to match the differential cross section from SCET${}_{\rm I}$ onto SCET${}_{\rm II}$. For clarity we will treat
the $^1S_0^{(8)}$ contribution explicitly and state the final result for the $^3P_J^{(8)}$ channel. The SCET${}_{\rm I}$ differential
cross section is
\bea\label{scetIcs}
2 E_\psi \frac{d\sigma}{d^3 p_\psi} &=&  \int  \, d^3 \bomega_{1} \int  \, d^3 \bomega_{2} 
\frac{-C^\dagger_{\beta\mu} C^\mu_\alpha}{32 \pi^3 s}
\sum_{X_n, X_s} (2\pi)^4 \delta^{(4)}(p_\gamma+p_P -p_\psi - p_s - p_n)
 \nn \\
&&\times \frac{1}{2} \sum_{\textrm{ spin}}
\langle p_P | J^{\beta\dagger}(\bomega_{1}, 0) | \jpsi + X_s +X_n \rangle
\langle \jpsi + X_s + X_n | J^\alpha(\bomega_{2} , 0) | p_P \rangle .
\eea
where $p_n$ is the total momentum of the final state collinear particles denoted by $X_n$, and $p_s$ is the total momentum of the final
state ultra-soft (usoft) particles, $X_s$. The momentum conserving delta function can be separated into light-cone coordinates:
\bea
\delta^{(4)}(p_\gamma+p_P -p_\psi - p_s - p_n) &=&  
2 \,  \delta\big( \sqrt{s} (1-z) - n\cdot p_n - n\cdot p_s \big) \, \delta\Big( \sqrt{s}\left( 1- \frac{m^2_\perp}{z s} \right) - \bn\cdot p_n \Big)
\nn \\
&\times& \delta^{(2)}(p_{\perp} + p_{n \perp} ) \,,
\eea
where $p_\perp$ is the $\jpsi$ transverse momentum. The usoft momentum is a subleading contribution in the last two terms on the
right-hand-side and has been dropped. As pointed out earlier, in the endpoint region  $|p_\perp| \ll M$ so from here on we let
$m^2_\perp \to M_\psi^2$. 

We factor the collinear degrees of freedom from the soft through a field redefinition which decouples usoft and  collinear fields in
the SCET${}_{\rm I}$ Lagrangian
\bea
B^\alpha_\perp(x) & \to & Y(x) B^{(0)\alpha}_\perp(x) Y^\dagger (x) \,.
\eea 
Here $Y(x)$ is a path ordered exponential of usoft gluon fields extending from $ - \infty$ to $x$. The above field
redefinition also shifts the collinear fields in the out state in such a way that $Y(x) \to \tilde{Y}(x)$ where $ \tilde{Y}(x)$ is a path
ordered exponential extending from $x$ to $\infty$~\cite{Arnesen:2005nk}
\begin{equation}
\tilde{Y}(x) = \textrm{P exp} \bigg( ig \int^\infty_0 ds \, n\cdot A_{us}(sn + x) \bigg) \,.
\end{equation}

As a consequence the matrix elements in Eq.~(\ref{scetIcs}) can be factored into separate collinear and usoft pieces. The cross 
section in its factored form is 
\bea
2 E_\psi \frac{d\sigma}{d^3 p_\psi} &=&  \int  \, d^3  \bomega_{1} \int  \, d^3 \bomega_{2} 
\frac{-C^\dagger_{\mu \beta}C^\mu_{\alpha}}{32 \pi^3 s}
\sum_{X_n, X_s} 2 (2\pi)^4 \delta\big( \sqrt{s} (1-z) - n\cdot p_n - n\cdot p_s \big) 
\nn \\
& \times &  \delta\Big( \frac{M_\psi^2}{z\sqrt{s}}  - \bar{\omega}_2 \Big)  \delta^{(2)}(p_{ \perp} - \omega_{2 \perp})
\nn \\
& \times & 2 \sum_{\textrm{ spin}}
\langle p_P |  \textrm{Tr} \big[ T^A B^\alpha_\perp(0) \big] \,  \delta^{(3)}({\vec{\cP}^\dagger - \bomega_2})  | X_n \rangle
\langle X_n |  \delta^{(3)}({\vec{\cP} - \bomega_1}) \,  \textrm{Tr} \big[ T^B B^\beta_\perp(0) \big] |p_P \rangle
\nn \\
& \times &
\langle 0 | \chi^\dagger_{-{\bf p}'} \tilde{Y} T^A \tilde{Y}^\dagger \psi_{{\bf p}'} (0) | \jpsi + X_s \rangle
\langle \jpsi + X_s | \psi^\dagger_{{\bf p}}  \tilde{Y}T^B \tilde{Y}^\dagger \chi_{-{\bf p}}(0) | 0 \rangle \,,
 \eea
where we have used conservation of the label momentum to replace the final state collinear momenta with $\omega$ and have 
dropped the $(0)$ superscript on the collinear fields. The momentum components $n\cdot p_n\sim O(\LQCD)$ are of the same
 size as the usoft momentum components. This remaining delta-function can be expressed as an integral over an exponential
\bea
(2 \pi) \, \delta\big( \sqrt{s} (1-z) - n\cdot p_n - n\cdot p_s \big) = 
\int \, \frac{dx^-}{2} \,  \textrm{exp} \Big[ \frac{i}{2} \big( \sqrt{s} (1-z) - n\cdot p_n - n\cdot p_s \big) x^- \Big] \,,
\eea
which can be pulled into the factored matrix elements where it shifts fields from the origin to the point $x^-$. The 
explicit dependence
on  $p_n$ and $p_s$ disappears and the sums over final collinear and usoft states can be performed using completeness. Using 
$d^3p_\psi/(2E_\psi)$ = $dz d^2p_{\perp}/(2 z)$ gives
\bea
\frac{d\sigma}{dz d^2p_{\perp}} &=& 
\frac{-1}{16 s z} \left( \frac{2 e e_c g_s}{M} \right)^2
 \int  \, d^3  \bomega_{+} \int \frac{dx^-}{2}  \,
e^{\frac{i}{2} \sqrt{s}(1-z)x^-} \, \delta \Big(   \frac{\bar{\omega}_+}{2} -\frac{M_\psi^2}{z\sqrt{s}} \Big) \, 
\delta^{(2)}\Big( p_{ \perp} -\frac{\omega_{+\perp}}{2}\Big)
\nn \\
& \times & \frac{1}{2} \sum_{\textrm{ spin}}
\langle p_P |   \textrm{Tr} \big[ B^\nu_\perp(x^-)  \,  \delta^{(3)}({\vec{\cP}_+ - \bomega_+}) B_{\perp \nu}(0) \big] |p_P \rangle
\nn \\
& \times &
\langle 0 | \chi^\dagger_{-{\bf p}'} \tilde{Y} T^A \tilde{Y}^\dagger \psi_{{\bf p}'} (x^-)  {\cal P}_\psi 
\psi^\dagger_{{\bf p}}  \tilde{Y} T^A \tilde{Y}^\dagger \chi_{-{\bf p}}(0) | 0 \rangle \,,
\eea
where we simplified the collinear matrix element by projecting onto a collinear operator in a color-singlet configuration. The expression 
above is further simplified by integrating over $\bomega_{+}$ and $p_{ \perp}$ 
\bea\label{scetIcs2}
\frac{d\sigma}{dz } &=& \frac{-1}{8 s z}  \left( \frac{2 e e_c g_s}{M} \right)^2 \int \frac{dx^-}{2}  \,
e^{\frac{i}{2} \sqrt{s}(1-z)x^-} \, 
\langle 0 | \chi^\dagger_{-{\bf p}'} \tilde{Y} T^A \tilde{Y}^\dagger \psi_{{\bf p}'} (x^-)  {\cal P}_\psi 
\psi^\dagger_{{\bf p}}  \tilde{Y} T^A \tilde{Y}^\dagger \chi_{-{\bf p}}(0) | 0 \rangle
\nn \\
& \times & \frac{1}{2} \sum_{\textrm{ spin}}
\langle p_P |   \textrm{Tr} \big[ B^\nu_\perp(x^-)  \,  \delta\bigg( \bnP- \frac{2 M_\psi^2 }{z\sqrt{s}} \bigg) 
B_{\perp \nu}(0) \big] |p_P \rangle \,, \nn \\
&=& \frac{-\pi}{2 s z}  \left( \frac{2 e e_c g_s}{M} \right)^2 \langle {\cal O}^{\jpsi}_8(^1S_0)  \rangle  M
 \int dk^+ S^{(8,{}^1S_0)}(-\sqrt{s}(1-z) + k^+) {\cal J}_P(k^+) \, .
\eea
The cross section is expressed as a convolution of a shape function,
$S^{(8,{}^1S_0)}$, and a jet function, ${\cal J}_P$, that are defined as follows:
 \bea\label{sdef}
 S^{(8,{}^1S_0)} (\ell^+) \equiv 
 \int \frac{d x^-}{4 \pi} \, e^{-\frac{i}{2} \ell^+ x^-}
\frac{ \langle 0 | \chi^\dagger_{-{\bf p}'} \tilde{Y} T^A \tilde{Y}^\dagger \psi_{{\bf p}'} (x^-) {\cal P}_\psi 
\psi^\dagger_{{\bf p}} \tilde{Y} T^A \tilde{Y}^\dagger \chi_{-{\bf p}}(0) |0\rangle}{ 2M \langle {\cal O}^{\jpsi}_8(^1S_0) \rangle} \, ,
\eea
\bea\label{jdef}
 {\cal J}_P(k^+) \equiv
 \int \frac{dy^-}{4\pi} e^{\frac{i}{2} k^+ y^-} \frac{1}{2} \sum_{\textrm{ spin}}
\langle p_P |   \textrm{Tr} \big[ B^\nu_\perp(y^-)  \,  \delta\bigg( \bnP- \frac{2 M_\psi^2 }{\sqrt{s}} \bigg) 
B_{\perp \nu}(0) \big] |p_P \rangle \,.
\eea
We have set $z \to 1$ inside the matrix element in the definition of the jet function 
since the matrix element is smooth as $z \to 1$ and leads to no large logs. We will also set $z\to 1$ in the 
prefactor of the cross section appearing in Eq.~(\ref{scetIcs2}). The shape function is normalized
so that $\int dk^+ S^{(8,{}^1S_0)}(k^+)=1$.

Taking moments of the cross section with respect to $z$, $\sigma_N \equiv \int_0^1 dz z^N d\sigma/dz$, and considering the large $N$ limit, we
will see below that $\sigma_N$ is the product of moments of the shape function and jet function. Large logs of $M/N$ are 
contained in moments of the shape function and large logs of $M/\sqrt{N}$ are contained in  the moments of ${\cal J}_P$, so the two scales 
are separated in Eq.~(\ref{scetIcs2}). However, the jet function still depends two scales, the perturbative scale, $M/\sqrt{N}$, as well as
the long distance scale, $\Lambda_{\rm QCD}$. Dependence on the latter scale arises because  the matrix element is taken  between proton
states. In this respect, the jet function that appears in Eq.~(\ref{scetIcs2}) is quite different from the perturbatively calculable jet
function that appears in the endpoint resummation of $e^+ + e^- \to \jpsi + X$ \cite{Fleming:2003gt}. 

Since the jet function, ${\cal J}_P$, contains both perturbative and nonperturbative scales, we can perform a further
factorization on this matrix element.  Because the external gluons in the proton have off-shellness  $O(\Lambda_{\rm QCD})$, the matrix
element should be evaluated in \SCETb rather than \SCETaa. The factorization can be thought of as  arising from matching the
nonlocal \SCETa operator in Eq.~(\ref{jdef}) onto local operators in \SCETbb. Intuitively we expect the result  to be a convolution
of short distance coefficient which is perturbatively calculable and  contains large logarithms of $M/\sqrt{N}$ with a parton
distribution function for the gluon in the proton. 

To demonstrate this  we define the variable, $\rho = M_\psi^2/s$, and take moments of ${\cal J}_P$ with respect to 
$\rho$:
\bea\label{Jmom}
\int_0^1 d\rho\,  \rho^N {\cal J}_P (k^+) &=& \int_0^1 d\rho \rho^N \int \frac{dy^-}{4\pi} e^{\frac{i}{2} k^+ y^-} 
\nn \\
& & \qquad \times \frac{1}{2} \sum_{\textrm{ spin}}
\langle p_P |   \textrm{Tr} \big[ B^\nu_\perp(y^-)  \,  \delta( \bnP- 2 \rho \sqrt{s} ) 
B_{\perp \nu}(0) \big] |p_P \rangle \nn \\
&=& \int\frac{dy^-}{4\pi}  e^{\frac{i}{2} k^+ y^-} \frac{1}{2} \sum_{\textrm{ spin}}
\langle p_P |   \textrm{Tr} \big[ B^\nu_\perp(y^-)  \, \frac{ \bnP^N}{(2 \sqrt{s})^{N+1}} 
B_{\perp \nu}(0) \big] |p_P \rangle\,.
\eea
Now we apply the operator product expansion (OPE) to the matrix element, which is justified when matching onto \SCETb because the momentum
$k^+\sim O(M/\sqrt{N})$ is parametrically larger than $\Lambda_{\rm QCD}$: 
\bea
\label{opeme}
\int_0^1 d\rho \, \rho^N {\cal J}_P (k^+) &=&  \int \frac{dy^-}{4\pi}  e^{\frac{i}{2} k^+ y^-} C(N,y^-) \frac{1}{2}  \sum_{\textrm{ spin}} 
\langle p_P |   \textrm{Tr} \big[ B^\nu_\perp  \, \frac{ \bnP^N}{(2 \sqrt{s})^{N+1}} 
B_{\perp \nu} \big] |p_P \rangle \nn \\
 &=&  - \frac{1}{4 s}\tilde C(N,k^+) \int_0^1 d \xi \xi^{N-1} f_{g/P}(\xi) \, ,
\eea
where 
\bea
\tilde C(N,k^+) = \sqrt{s}  \int\frac{dy^-}{4\pi}  e^{\frac{i}{2} k^+ y^-} C(N,y^-) \, ,
\eea 
is dimensionless.
Therefore the moments of the jet function are proportional to the moments of the structure function
for the gluon in the proton, whose definition in SCET is ~\cite{Bauer:2002nz}
\begin{equation}
\frac{1}{2} \sum_{\textrm{spin}} 
\langle p_{P} | 
\big[  \textrm{Tr}\big\{ B_\perp^{\nu}(0)  \delta(\bnP_+ - \bar{\omega}_+)  B_{\perp \nu}(0) \big\}  \big]
| p_{P} \rangle 
=-\frac{1}{2 \bar{\omega}_+}   f_{g/P}\Big(\frac{\bar{\omega}_+}{2\bn\cdot p_P}\Big)  \,.
\end{equation}

The result in Eq.~(\ref{opeme}) is easily seen to be equivalent to the following convolution:
\bea
{\cal J}_P (k^+) =  \frac{-1}{4 \rho s}\int_\rho^1 \frac{d\xi}{\xi} \, C_{II}\left(\frac{\rho}{\xi},k^+\right) f_{g/P}(\xi)\,,
\eea
where the function $C_{II}(\rho/\xi , k^+)$ can be obtained from the coefficients in the OPE, $\tilde C(N,k^+)$,
by inverse Mellin transform. The resulting expression for the cross section is 
\bea \label{cs}
\frac{d \sigma}{dz} = \sigma_0 \rho 
\int dk^+ S^{(8,{}^1S_0)}(-\sqrt{s}(1-z) +k^+) \int_\rho^1 \frac{d\xi}{\xi} \, C_{II}\left(\frac{\rho}{\xi},k^+\right) f_{g/P}(\xi) \, ,
\eea 
where 
\bea
\sigma_0 = \frac{\pi^3 \alpha \alpha_s e_c^2}{4 m_c^5}  \langle {\cal O}^{\jpsi}_8(^1S_0) \rangle \nn \, ,
\eea 
and we have set $\rho= 4m_c^2/s$. The factorized form  for the color-octet P-wave contribution
to the cross section is obtained by making the replacement
\bea
 \langle {\cal O}^{\jpsi}_8(^1S_0) \rangle \to \frac{7}{m_c^2} \langle {\cal O}^{\jpsi}_8(^3P_0) \rangle
\, ,
\eea
and replacing $S^{(8,{}^1S_0)}$ with $S^{(8,{}^3P_0)}$. 
The shape functions in each channel are normalized the same way.
Eq.~(\ref{cs}) is modified once higher order corrections to the matching coefficients in
Eq.~(\ref{match}) are included. We parametrize these corrections as follows:
\bea\label{match_ho}
C^{\mu \alpha} (^1S_0^{(8)})&=& C_V(\bar \mu)\,\frac{- 2 e e_c g_s(M)}{M} \epsilon_\perp^{\mu \alpha}
\eea
and similarly for $^3P_J^{(8)}$ channels. Including these corrections gives the final form of the factorization theorem:
\bea \label{csc}
\frac{d \sigma}{dz} = \sigma_0 \rho |C_V(\bar \mu)|^2 \,
\int dk^+ S^{(8,{}^1S_0)}(-\sqrt{s}(1-z) +k^+) \int_\rho^1 \frac{d\xi}{\xi} \, C_{II}\left(\frac{\rho}{\xi},k^+\right) f_{g/P}(\xi) \, .
\eea 
This factorization theorem for endpoint photoproduction of $J/\psi$   is the main result of this paper.
For the remainder of this section we set $C_V(\bar \mu) = 1$. In section V, where  large 
logarithms are resummed, an important step  is evolving $C_V(\bar \mu)$ from the scale $\bar \mu = M$, where QCD is matched onto \SCETaa,
to the scale $\bar \mu = M/\sqrt{\bar N}$, where \SCETa is matched onto \SCETbb.

Our result should reproduce previous results for photoproduction in the appropriate limits.
To lowest order in $\alpha_s$ the coefficients in the OPE are $C(N,y^-) =1$, which is equivalent to
\bea
C_{II}\left(\frac{\rho}{\xi},k^+\right)= \sqrt{s} \, \delta(k^+)\,\delta\left(1- \frac{\rho}{\xi}\right) \, .
\eea
Inserting this into Eq.~(\ref{cs}) yields
\bea 
\frac{d \sigma}{dz} =\frac{\pi^3 \alpha \alpha_s e_c^2}{s m_c^3} \langle {\cal O}^{\jpsi}_8(^1S_0) \rangle
\, \sqrt{s} \, S^{(8,{}^1S_0)}(-\sqrt{s}(1-z) ) \, f_{g/P}(\rho) \, .
\eea 
To lowest order in $v^2$, $\sqrt{s} \, S^{(8,{}^1S_0)}(-\sqrt{s}(1-z) ) 
\rightarrow \delta(1-z)$ which is easily seen to reproduce the tree level calculation of 
Ref.~\cite{Amundson:1996ik}. 

Finally, it is useful to derive an expression for the moments of the cross section. Defining
$k^+ = \sqrt{s}(u-z)$, the normalized differential cross section is 
\bea
\frac{1}{\sigma_0}\frac{d \sigma}{dz} =   \rho \int_z^1 du \, \hat{S}^{(8,{}^1S_0)}(u) 
\int_\rho^1 \frac{d\xi}{\xi} \, C_{II}\left(\frac{\rho}{\xi},u-z\right) f_{g/P}(\xi) \, ,
\eea
where $\hat{S}^{(8,{}^1S_0)}(u) = \sqrt{s}S^{(8,{}^1S_0)}(-\sqrt{s}(1-u))$ and we have rescaled  second argument
of $C_{II}$. It is now straightforward to 
take moments of the cross section and show that, for large $N$, the normalized moments of the 
cross section factorize:
\bea\label{csmom}
\frac{\sigma_N}{\sigma_0} \equiv \frac{1}{\sigma_0}\int_0^1 dz \, z^N \frac{d\sigma}{dz} = 
 \rho \tilde{S}^{(8,{}^1S_0)}(N) \int_\rho^1 \frac{d\xi}{\xi} \, \hat{C}_{II}\left(\frac{\rho}{\xi},N \right) f_{g/P}(\xi) \, ,
\eea
where 
\bea 
\tilde{S}^{(8,{}^1S_0)}(N) &=& \int_0^1 du \, u^N \hat{S}^{(8,{}^1S_0)}(u)\,, \nn \\
\hat{C}_{II}\left(\frac{\rho}{\xi},N \right) &=& \int_0^1 dz \, z^N C_{II}\left(\frac{\rho}{\xi},1-z\right) \, .
\eea
We will show in Sec.~\ref{mr} that the moments of the QCD cross section  at lowest nontrivial order 
factorize in a manner consistent with Eq.~(\ref{csmom}), providing additional evidence
for the factorization theorem.

\section{Extracting Large Logs at NLO}

A calculation of the leading color-octet contribution 
to the forward cross section for $J/\psi$ photoproduction appeared in Ref.~\cite{Amundson:1996ik}.
The NLO calculation of the color-octet contribution to $d \sigma/dz, z \neq 1$, 
was carried out in Refs.~\cite{Cacciari:1996dg, Kniehl:1998qy}, but these papers presented only numerical results. 
Refs.~\cite{Ko:1996xw,Maltoni:1997pt} presented analytic results for the total partonic cross section.
In this section we obtain analytic expressions for the leading logarithmically enhanced corrections 
to the color-octet differential cross section in NRQCD. By logarithmically enhanced we mean 
corrections of the form:
\bea
\frac{1}{\sigma_0}\frac{d \hat \sigma[\gamma + g \to J/\psi+X]}{dz} \propto 
\alpha_s \left(\frac{{\rm ln}(1-z)}{1-z}\right)_+, \ \alpha_s \left(\frac{1}{1-z}\right)_+ \, .
\label{largelog}
\eea
When moments of the cross section are taken, $\int_0^1 dz z^N d\sigma/dz$, the distributions
in Eq.~(\ref{largelog}) give rise to terms $\propto \alpha_s {\rm ln}^2 N$ and $\alpha_s {\rm ln} N$,
respectively. These are the logarithmically enhanced terms we wish to resum to all orders in 
perturbation theory. Terms in the differential cross section less singular in $1-z$
give contributions to the moments which are suppressed by powers of $N$, e.g.
$\int_0^1 dz z^N {\rm ln}(1-z) =  {\rm ln}\bar{N}/N +O(1/N^2)$, where $\bar{N}= N e^{\gamma_E}$, and will be ignored.

The analytic result for the logarithmically enhanced corrections is helpful  for checking our factorization theorem for the  resummed cross
section. Furthermore, in our  derivation of the factorization theorem, QCD currents are first matched onto \SCETa  operators, then the
cross section  in \SCETa  is matched onto  \SCETbb. In both matching calculations, we wish to verify that no large logarithms appear in the
matching coefficients so we are sure that large logarithms are properly resummed using the RGEs. The condition that the large logs cancel
in the matching from \SCETa and \SCETb determines the scale at which \SCETa is matched onto \SCETbb.

The hadronic cross section is obtained by convoluting the partonic cross section 
with a parton distribution function (pdf):
\bea\label{conv}
\frac{d \sigma[\gamma + P \to J/\psi + X]}{d z} = \sum_i \int_\rho^z dx \, \frac{\rho}{x^2} 
f_{i/P}\left(\frac{\rho}{x}\right) \frac{d \hat \sigma[\gamma + i \to J/\psi + X]}{dz} \, .
\eea
Here $\rho = M^2/s$ and $x = M^2/\hat{s}$, where $\hat{s}$ is the center of mass energy squared in  the partonic collision. There is a sum
over parton species, $i$, and $f_{i/P}$ is the pdf for the proton. For the rest of the paper we will restrict ourselves to the dominant
production  process which is initiated by a gluon. In the HERA collider $J/\psi$ photoproduction experiments $\sqrt{s} \sim 100 \, {\rm
GeV}$ so $\rho \approx 10^{-3}$. Though we cannot set $\rho$ to zero inside the argument of the pdf, we can set $\rho = 0 $ in the limit of
the $x$ integration since the logarithmically enhanced terms come from the opposite end of the integral. Calculation of the partonic
cross section is simplified by making this  approximation, so we will use the standard distributional identity
\bea\label{stan}
(1-x)^{-1-\eps} =  -\frac{1}{\eps} \delta(1-x) + \left( \frac{1}{1-x}\right)_+ 
-\eps\left(\frac{{\rm ln}(1-x)}{1-x}\right)_+ + O(\eps^2)\,,
\eea
instead of \cite{Maltoni:1997pt} 
\bea\label{rho}
(1-x)^{-1-\eps} =  -\frac{\beta^{-2 \eps}}{\eps} \delta(1-x) + \left( \frac{1}{1-x}\right)_\rho
-\eps\left(\frac{{\rm ln}(1-x)}{1-x}\right)_\rho + O(\eps^2)\,,
\eea
where $\beta =\sqrt{1-\rho}$ and the $\rho$-distributions are defined by
\bea
\int_\rho^1 dx  \, [t(x)]_\rho \, f(x) = \int_\rho^1 dx \, t(x) \, [f(x)-f(1)] \, . 
\eea
By using Eq.~(\ref{stan}) rather than Eq.~(\ref{rho}) we omit corrections proportional to $\log \beta \sim 10^{-3}$ which 
can be safely neglected.

\begin{figure}
\begin{center}
\includegraphics[width=3in]{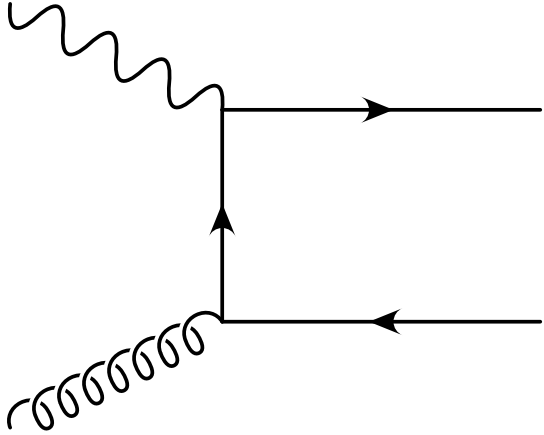}
\caption{\it  One of two leading order QCD diagram  for $\gamma g\to c\bar{c}^{(8)}(^{2S+1}L_J)$. The other Feynman graph is obtained by reversing the direction of the fermion line}
\label{ggcclo}
\end{center}
\end{figure}
\begin{figure}
\begin{center}
\includegraphics[width=3in]{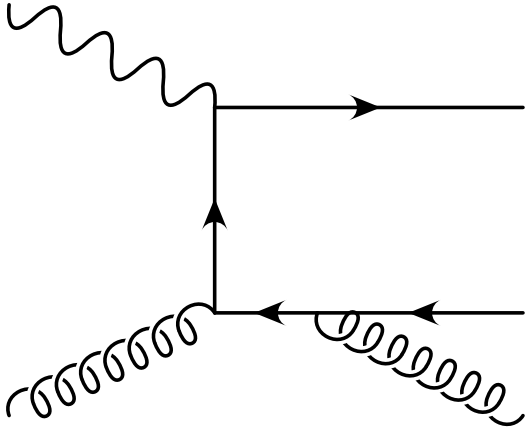}
\caption{\it  An example of a NLO QCD diagram contributing to photoproduction. } 
\label{ggccnlo}
\end{center}
\end{figure}

The leading order contribution to the production of $J/\psi$ via color-octet mechanisms 
is the two-to-one process, $\gamma g \to c \bar{c}^{(8)}(^{2S+1}L_J)$, where one of the two 
Feynman diagrams is depicted in Fig.~\ref{ggcclo}. 
This gives a contribution to the partonic cross section~\cite{Amundson:1996ik}
\bea 
\label{partcs}
\frac{d \hat \sigma^{\rm LO }}{dz} &=& \frac{\pi^3 \alpha_s \alpha e_c^2}{4 m_c^5} \,
\left[ \langle {\cal O}^{J/\psi}_8({}^1S_0)\rangle + \frac{7}{m_c^2} 
\langle {\cal O}^{J/\psi}_8({}^3P_0)\rangle\, \right] \delta(1-x) \, \delta(1-z)  \nn \\
&\equiv& \sigma_0 \, \delta(1-x)\, \delta(1-z)  \,.
\eea 
An example of a next-to-leading order real emission diagram 
that contributes to the photoproduction cross section for $z \neq 1$ is shown in Fig.~\ref{ggccnlo}.
It is these graphs that contribute the large logarithms at NLO we wish to resum. 
Infrared (IR) divergences in these graphs will be regulated using dimensional regularization 
and the IR divergences are cancelled by virtual corrections to the LO graphs, Fig.~\ref{ggcclo}, 
or by the parton distribution function. 

The partonic cross section obtained from summing all the NLO real emission diagrams in the color-octet
${}^1S_0$ channel is  
\bea
\frac{d \hat{\sigma}^{\rm real}}{d z}  = \sigma_0 \frac{C_A \alpha_s}{\pi} 
\left(\frac{4 \pi \mu^2}{M^2}\right)^\epsilon \frac{x^\epsilon (1-z)^{-\epsilon} (z-x)^{- \epsilon}}{\Gamma[1-\epsilon]}{\cal M}(x,z) \, ,
\eea
where 
\bea \label{Mxz}
{\cal M}(x,z) = \frac{x^2(z-x)\left[(1-z+z^2)^2+x (x-z) (x(x-z)+2 z^2)\right]}{(1-x)^2 (1+x-z)^2(1-z)z^2} \, .
\eea
The phase space is in $D=4 -2 \epsilon$ dimensions. The  logarithimically enhanced terms come from the limit 
$z \to 1$. Expanding $M(x,z)$ about $z=1$ we find
\bea\label{zexp}
{\cal M}(x,z) = \frac{(1-x+x^2)^2}{(1-x)(1-z)}+ \frac{-2+5x-6x^2+5x^3-6x^4+5x^5-2x^6}{x(1-x)^2} 
+ O(1-z) \, .
\eea  
Terms $O(1-z)$ and higher do not give logarithmically enhanced contributions, even after performing the $x$ integration, and therefore can be
dropped. Furthermore, 
\bea
\int^z dx \frac{1}{(1-x)^2} \propto \frac{1}{1-z} \qquad \int^z dx \frac{1}{(1-x)} \propto {\rm ln}(1-z)\,,
\eea
so terms proportional to $1/(1-x)^2$ give logarithimically enhanced corrections while terms 
proportional to $1/(1-x)$ can be dropped. Thus the second term on the right hand side of
 Eq.~(\ref{zexp}) can be expanded around $x=1$,
\bea\label{ll}
\frac{d \hat{\sigma}^{\rm real}}{d z} &=& \sigma_0 \frac{C_A \alpha_s}{\pi} 
\left(\frac{4 \pi \mu^2}{M^2}\right)^\epsilon \frac{x^\epsilon (1-z)^{-\epsilon} (z-x)^{- \epsilon}}{\Gamma[1-\epsilon]}
\left[ \frac{(1-x+x^2)^2}{(1-x)(1-z)} - \frac{1}{(1-x)^2} + ...\right]\, \nn \\
&=& \sigma_0\frac{\alpha_s}{2\pi} 
\left(\frac{4 \pi \mu^2}{M^2}\right)^\epsilon \frac{x^\epsilon (1-z)^{-\epsilon} (z-x)^{- \epsilon}}{\Gamma[1-\epsilon]} 
\left[\frac{x P_{gg}(x)}{1-z} - \frac{2 C_A}{(1-x)^2} + ...\right]\, ,
\eea
where the ellipsis represents terms that do not give logarithmically enhanced corrections
and $P_{gg}(x)$ is the related to the real emission part of the gluon splitting function:
\bea
P_{gg}(x) = 2 C_A \left[\frac{x}{1-x} +  \frac{1-x}{x}+x(1-x)\right] \nn \, .
\eea
The gluon splitting function, ${\cal P}_{gg}(x)$, will appear in what follows and is defined by: 
\bea
{\cal P}_{gg}(x) &=& {\overline P}_{gg}(x) + b_0 \, \delta(1-x)\,, \nn \\
{\overline P}_{gg}(x) &=& 2 C_A \left[\frac{x}{(1-x)_+} +  \frac{1-x}{x}+x(1-x)\right] \,, \nn \\
b_0 & = & \frac{11}{6}C_A -\frac{2}{3} T_F n_f \, .
\eea
In Eq.~(\ref{ll}) the first term in square brackets contains the collinear divergence, as is easily seen by noting that $1-z = (1-x)(1-\cos\theta)/2$, where $\cos \theta$ is the
angle between initial and final state gluon momentum in the parton center of mass frame. 

Though we have derived Eq.~(\ref{ll}) for the $^{1}S_0$ channel, the result is in fact universal 
and holds for the color-octet channels $^{1}S_{0}, {}^{3}P_{0,2}$ regardless of the angular momentum quantum numbers of the 
final state $c \bar c$. QCD factorization theorems guarantee that the term with the
$1/(1-z)$ pole in the real emission cross section has the same form for all color-octet production 
channels. The universality of the double pole, $1/(1-x)^2$, is verified by calculating QCD diagrams 
in the soft gluon approximation, which is valid in the limit $x \to 1$~\cite{Maltoni:1997pt}.

To obtain  the logarithmically enhanced corrections we need to perform the integration over $x$. At first sight, deriving an analytic
expression appears to be difficult  since the pdf contains $x$ dependence which is not known in closed form. However, it is possible
to extract the leading corrections analytically  using the  following  distributional identities, derived in Appendix A:
\bea\label{tr1}
(1-z)^{-1-\eps}\int_0^z d x \frac{(z-x)^{-\eps}}{1-x} g(x) &=& \\
&& \hspace{-2.0 in} \delta(1-z)\left[\left(\frac{1}{2\eps^2}-\frac{\pi^2}{12}\right) g(1)
- \frac{1}{\eps}\int_0^1 dx \left(\frac{1}{1-x}\right)_+ g(x)
+ \int_0^1 dx\left(\frac{{\rm ln}(1-x)}{1-x} \right)_+ g(x) \right]  \nn \\
&& \hspace{-2.0 in}
 +\left(\frac{1}{1-z}\right)_+ \int_0^1 d x \left(\frac{1}{1-x}\right)_+ g(x)
 -g(1)\left(\frac{{\rm ln}(1-z)}{1-z} \right)_+  \, ,\nn \\
 && \nn \\
\int_0^z d x \frac{(z-x)^{-\eps}(1-z)^{-\eps}}{(1-x)^2} g(x) &=& 
\left[-\frac{1}{2\eps} \delta(1-z) + \left(\frac{1}{1-z}\right)_+\right] g(1) \, , \label{tr2}
\eea
where $g(x)$ is arbitrary. On the right hand side we only keep terms proportional to singular distributions in $1-z$. This
includes the singular distributions in  Eq.~(\ref{largelog}) which contribute to the large logarithms in moment space as well as terms
proportional to $\delta(1-z)$. These terms contain IR divergences which cancel against virtual graphs or are absorbed into the pdf. 

The NLO result for the part of the partonic real emission contribution to $d\sigma/dz$
that is singular as $z \to 1$ is 
\bea\label{QCDcsNLO1}
\frac{d \hat \sigma^{\rm real}}{d z} &=& \sigma_0 \frac{C_A \alpha_s}{\pi} \times\\
&& \hspace{-2ex}
\left( \delta(1-z) \left\{ \delta(1-x) \left[\frac{1}{2\eps^2} + \frac{1}{2\eps} + \frac{1}{2\eps}{\rm ln}
\left(\frac{\bar \mu^2}{M^2}\right) + \frac{1}{2}{\rm ln}\left(\frac{\bar \mu^2}{M^2}\right)
 +\frac{1}{4} {\rm ln}^2\left(\frac{\bar \mu^2}{M^2}\right) -\frac{\pi^2}{8}\right]  
\right. \right. \nn \\
&&\left. +  (1-x+x^2)^2 
\left[\left(-\frac{1}{\eps}- {\rm ln}\left(\frac{\bar \mu^2 x}{M^2}\right) \right)
\left(\frac{1}{1-x}\right)_+  + \left(\frac{{\rm ln}(1-x)}{1-x}\right)_+ \right] 
 \right\} \nn \\
&&+ \left.  \left(\frac{1}{1-z}\right)_+ \left[-\delta(1-x) +  (1-x+x^2)^2
\left(\frac{1}{1-x}\right)_+ \right]  -\delta(1-x) \left(\frac{{\rm ln}(1-z)}{1-z}\right)_+ \right) \, ,\nn
\eea
where we define $\bar \mu^2 = 4 \pi \mu^2 e^{-\gamma_E}$.
If we take moments of this cross section with respect to the variable $z$ we find 
\bea\label{QCDcsNLO2}
\frac{\hat \sigma^{\rm real}_N}{\sigma_0} &=& \delta(1-x) \frac{C_A \alpha_s}{\pi} \times \\
&& \left[ \frac{1}{2 \eps^2} + \frac{1}{2 \eps} + \frac{1}{2\eps} {\rm ln}
\left(\frac{\bar \mu^2}{M^2}\right) +\frac{1}{4}{\rm ln}^2\left(\frac{\bar \mu^2}{M^2}\right) 
+\frac{1}{2}{\rm ln}\left(\frac{\bar \mu^2}{M^2}\right) 
+{\rm ln}(\bar N) -\frac{1}{2}{\rm ln}^2(\bar N) -\frac{\pi^2}{8} \right] \nn\\
&& +  \frac{\alpha_s}{2\pi} x {\overline P}_{gg}(x) 
\left[-\frac{1}{\eps}- {\rm ln}\left(\frac{\bar \mu^2 x \bar N}{M^2}\right)\right]
+  \frac{\alpha_s}{2\pi} x (1-x) P_{gg}(x) \left(\frac{{\rm ln}(1-x)}{1-x}\right)_+ \,.
\nn
\eea
This expression for the partonic QCD cross section is needed for the matching 
calculations in the next section.

\section{matching}
\label{mr}

In this section, we match QCD onto \SCETa at the scale $\bar \mu = M$ and then \SCETa onto \SCETb  at
the scale $\bar \mu = M/\sqrt{\bar N}$.  At the first stage we must verify that \SCETa reproduces the IR behavior of QCD, i.e.  the \SCETa
real emission diagrams must reproduce the terms in Eq.~(\ref{ll}) which dominate as $z \to 1 $. At the second stage of matching, the \SCETa
cross section is matched on the \SCETb cross section, which takes on the factorized form of Eq.~(\ref{cs}). These calculations are
simplest  if dimensional regularization is used to regulate both ultraviolet (UV) and IR divergences. The matching calculations are important for
verifying the factorization theorem  and  determining the QCD-\SCETa and \SCETa-\SCETb matching  scales, which give the boundary
conditions for renormalization group evolution.

\subsection{Matching QCD onto \SCETa}

In the first step we match the QCD amplitude onto the \SCETa current at one loop. The one-loop QCD result
for the virtual correction can be found in Ref.~\cite{Maltoni:1997pt},
\bea\label{oneloopQCD}
 \frac{d\hat \sigma}{dz} &=& (\sigma_0 +\sigma^{(V)})\delta(1-x)\delta(1-z)\,, \\
\sigma^{(V)} &=& \sigma_0 \frac{\alpha_s}{2 \pi} f_\epsilon(M^2) \left\{ \frac{b_0}{\epsilon_\textrm{UV}} 
+\left(C_F-\frac{1}{2}C_A\right)\frac{\pi^2}{v}
-C_A\left( \frac{1}{\epsilon^2_\textrm{IR} }+ \frac{17}{6 \epsilon_\textrm{IR} }\right) + \frac{2}{3 \epsilon_\textrm{IR}} n_f T_F + 
2 D^{[8]}_{\cal O} 
\right\} \,,\nn
\eea
where $C_F=4/3$, $C_A=3$, $T_F=1/2$ are SU(3) group theory factors, $n_f$ is the number of light flavors, $v$ is the relative velocity of
the $c$ and $\bar{c}$ in the rest frame of the $c\bar{c}$ pair, and 
\bea
 f_\epsilon(M^2) & = & \left( \frac{4 \pi \mu^2}{M^2}\right)^\epsilon \Gamma(1+\epsilon )\,,\nn \\
 D^{[8]}_{^1S_0} &=& C_F\left( -5 + \frac{\pi^2}{4} \right) + C_A \left( \frac{3}{2} + \frac{\pi^2}{12} \right) \,,\nn \\
 D^{[8]}_{^3P_0} &=& C_F\left( -\frac{7}{3} + \frac{\pi^2}{4} \right) + C_A \left( \frac{1}{2} + \frac{\pi^2}{12} \right)\,, \nn \\
 D^{[8]}_{^3P_2} &=& -4 C_F+ C_A \left( \frac{3}{4}+ \frac{\log 2}{2} + \frac{\pi^2}{3} \right) \,.
 \eea
The UV divergence in Eq.~(\ref{oneloopQCD}) is removed by QCD coupling constant renormalization. When computing the matching coefficient
to this order,  one must include one-loop NRQCD virtual corrections to the matrix element of the NRQCD color-octet production operators.
The NRQCD corrections reproduce the Coulomb correction $\propto \pi^2/v$ in Eq.~(\ref{oneloopQCD}) and therefore this term does not appear
in the matching coefficient~\cite{Petrelli:1997ge}.

In \SCETa loop integrals are scaleless in dimensional regularization and therefore are zero. We must also 
subtract the contribution from the EFT counterterm which was calculated in Ref.~\cite{Bauer:2001rh}:
 \bea\label{cntrtrm}
 Z_{\cal O} -1 = \frac{\alpha_s}{4 \pi} \left[ C_A \left( \frac{1}{\epsilon^2} 
+\frac{1}{\epsilon}\log\left( \frac{\bar \mu^2}{M^2}\right) + 
 \frac{17}{6 \epsilon} \right) - \frac{2}{3 \epsilon} n_f T_F \right] \, .
 \eea
 Thus the \SCETa one loop result for the virtual cross section is
 \bea
 \frac{d \hat \sigma}{dz}= \sigma_0 |C_V (\bar \mu)|^2 \,\delta(1-z)\, \left(1 + \frac{\alpha_s}{2 \pi}\left\{ -C_A \left[ \frac{1}{\epsilon^2} +\frac{1}{\epsilon}\log\left( \frac{\bar \mu^2}{M^2}\right) + 
 \frac{17}{6 \epsilon} \right] +  \frac{2}{3 \epsilon} n_f T_F \right\} \right)\,.
 \eea
Taking the difference between this and Eq.~(\ref{oneloopQCD}) (after subtracting the UV divergence
and dropping the Coulomb correction)  we obtain the one loop matching coefficient
 \bea
 C_V(\bar \mu) = 1 +\frac{\alpha_s}{4 \pi} \left[ -C_A \left( \frac{1}{2} \log^2 \frac{\bar \mu^2}{M^2}+ 
 \frac{17}{6} \log \frac{\bar \mu^2}{M^2} +\frac{\pi^2}{12}\right) + \frac{2}{3} n_f T_F
 \log \frac{\bar \mu^2}{M^2} + 2 D^{[8]}_{\cal O} \right] 
\,.
 \eea
The logarithms are minimized for $\bar \mu = M$, which fixes the QCD-\SCETa matching scale.
 
\subsection{The NLO \SCETa differential cross section and matching onto \SCETb}
 
In the second step we match the \SCETa differential cross section onto the \SCETb differential cross section which has the factored form
given in Eq.~(\ref{cs}).  The calculation of the \SCETa differential cross section at NLO also allows us to confirm that the EFT reproduces
the parts of the NLO QCD calculation that are singular as $z\to1$. This is an  important check on the validity of the EFT. 

\begin{figure}
\begin{center}
\includegraphics[width=4in]{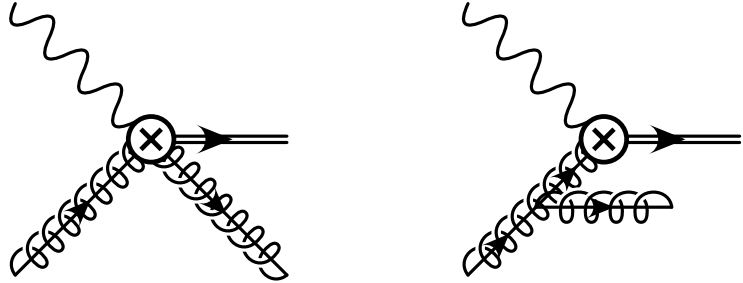}
\caption{\it Feynman graphs for the real emission of a collinear gluon.}
\label{SCETRealColl}
\end{center}
\end{figure}

\begin{figure}
\begin{center}
\includegraphics[width=4in]{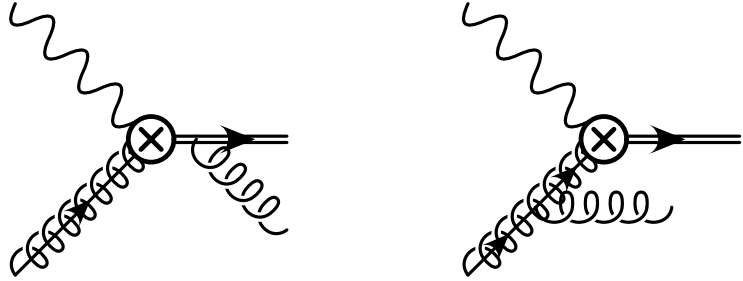}
\caption{\it Feynman graphs for the real emission of a usoft gluon.}
\label{SCETRealSoft}
\end{center}
\end{figure}

The SCET diagrams for the real emission come from 
both collinear and usoft graphs. These are shown in Figs.~(\ref{SCETRealColl}) and (\ref{SCETRealSoft}), respectively. 
The result of the usoft graphs is the
same as making the eikonal approximation in the full QCD diagram~\cite{Maltoni:1997pt}:
\bea\label{soft}
\frac{d \hat \sigma}{dz}^{\rm soft} = \sigma_0 \frac{C_A \alpha_s}{\pi} 
\frac{x^\epsilon (1-z)^{-\eps} (z-x)^{-\eps}}{\Gamma[1-\eps]}\left[ \frac{x^2}{(1-z)(1-x)}-\frac{x^2}{(1-x)^2}\right] \,.
\eea
Evaluation of the collinear graphs is more subtle. A naive evaluation employing the Feynman rules of 
Ref.~\cite{Fleming:2003gt} yields 
\bea \label{col}
\frac{d \hat \sigma}{dz}^{\rm col} = \sigma_0 \frac{C_A \alpha_s}{\pi}
\frac{x^\epsilon (1-z)^{-\eps} (z-x)^{-\eps}}{\Gamma[1-\eps]}
\left[ \frac{x^2(1+ x-z +(x-z)^2)^2}{(1-z)(z-x)(1+x-z)^2}\right] \,.
\eea
Note that the naive collinear contribution to the cross section has pole in $z-x$
that is not present in the full QCD calculation in Eq.~(\ref{Mxz}). Combining the cross sections
in Eq.~(\ref{soft}) and Eq.~(\ref{col}) then expanding about $z=1$ as was done earlier for ${\cal M}(x,z)$ 
does not yield the two leading terms in Eq.~(\ref{ll}). 

The origin of this problem is double counting of modes in certain corners of phase space 
where modes can be considered either collinear or usoft~\cite{Manohar:2006nz}. Collinear particles in SCET 
have momentum scaling as $Q(\lambda^2,1,\lambda)$. For the bulk of the phase space integration,
this scaling is satisfied but when $x, z \to 1$ simultaneously it is not. To see this note that the 
momentum of the final state gluon in the real emission diagrams is 
\bea
k^\mu &=& \frac{1}{2}\sqrt{\hat s}(1-z) \bn^\mu  -\frac{1}{2}\sqrt{\hat s}(x-z) n^\mu + k_\perp^\mu\,, \nn \\
k_\perp^2 &=& -{\hat s}(x-z)(1-z) \, .
\eea
 For $1-z \sim O(\lambda^2)$ and $x-z \sim O(1)$ 
we see that the final state gluon momentum scales as 
\bea 
(n\cdot k, \bn \cdot k, k_\perp) \sim \sqrt{\hat s} (\lambda^2, 1, \lambda) \, , \nn
\eea
appropriate for a collinear particle. However, the phase space integral over $x$ ranges from (nearly) zero
to $z$, so there is a corner of phase space in which $x- z \sim O(\lambda^2)$ rather than $O(1)$  and therefore
\bea 
(n\cdot k, \bn \cdot k, k_\perp) \sim \sqrt{\hat s} (\lambda^2,\lambda^2 , \lambda^2 ) \, .\nn
\eea
In this regime the scaling is appropriate for a usoft particle. 

In the SCET label formalism, double counting arises from a naive evaluation of the sums over collinear label
momenta appearing in loops and phase space integrations. The mode with vanishing label momentum, the so-called ``zero-bin", must be excluded
from these sums since a mode with vanishing collinear label momentum should be considered soft. As discussed in Appendix B, 
the standard trick for converting sums over label momenta with integrations over residual momenta to obtain ordinary
loop  and phase space integrals actually includes the zero-bin mode, and therefore certain zero-bin subtractions
must be made to correctly evaluate collinear diagrams. 

In our problem, the zero-bin subtraction can be implemented by including  additional  diagrams that are identical to the  
collinear diagrams except that the momentum of the final state gluon is treated as usoft and the appropriate 
approximations are made.  This graph  gives 
\bea \label{zb}
\frac{d \sigma}{dz}^{\rm zb} = \sigma_0 \frac{C_A \alpha_s}{\pi}
\frac{x^\epsilon (1-z)^{-\eps} (z-x)^{-\eps}}{\Gamma[1-\eps]}
\left[ \frac{-x^2 }{(1-z)(z-x)}\right] \, .
\eea
The minus sign reflects the fact that we must subtract the zero-bin contribution from the naive evaluation of the
collinear  contribution. Combining the usoft, collinear, and zero-bin subtraction yields
\bea
\frac{d \sigma}{dz} &=& \sigma_0 \frac{C_A \alpha_s}{\pi}
\frac{x^\epsilon (1-z)^{-\eps} (z-x)^{-\eps}}{\Gamma[1-\eps]} \times \nn \\
&&\left[ \frac{x^2 }{(1-z)(z-x)}\left( \frac{ (1+x-z+(x-z)^2)^2}{(1+x-z)^2} - 1\right)   
+ \frac{x^2}{(1-z)(1-x)} -\frac{x^2}{(1-x)^2} \right] \nn \\
&=&\sigma_0 \frac{C_A \alpha_s}{\pi} \frac{x^\epsilon (1-z)^{-\eps} (z-x)^{-\eps}}{\Gamma[1-\eps]}
\left[ \frac{(1+x-x^2)^2}{(1-z)(1-x)} -\frac{1}{(1-x)^2} + ...\right] \, ,
\eea
where the ellipsis represents terms which do not contribute logartihmically enhanced corrections.
We see that once the zero-bin contribution is subtracted, the spurious pole at $x=z$ is removed and the
leading behavior of the QCD cross section is reproduced by \SCETaa, as expected.

To get the \SCETa differential cross section to this order we must add the virtual contributions.
Loops are scaleless and therefore vanish in dimensional regularization, but the contribution
from tree level graphs with the counterterm in Eq.~(\ref{cntrtrm}) must also be included. We 
will match the moments of the  partonic cross sections rather than the partonic differential
cross sections. Combining the \SCETa real emission contribution with the \SCETa counterterm contribution, 
we obtain
\bea\label{scetImom}
\frac{\hat \sigma^I_N}{\sigma_0} &=&
-\frac{\alpha_s}{2 \pi} x {\cal P}_{gg}(x) \frac{1}{\eps}
\nn \\
&& + \delta(1-x) \frac{C_A \alpha_s}{\pi} 
 \left[  \frac{1}{4}{\rm ln}^2\left(\frac{\bar \mu^2}{M^2}\right) 
+\frac{1}{2}{\rm ln}\left(\frac{\bar \mu^2}{M^2}\right) 
+{\rm ln}(\bar N) -\frac{1}{2}{\rm ln}^2(\bar N) -\frac{\pi^2}{8} \right] \nn\\
&& - \frac{\alpha_s}{2\pi} x {\overline P}_{gg}(x) 
{\rm ln}\left(\frac{\bar \mu^2 x \bar N}{M^2}\right)
+  \frac{\alpha_s}{2\pi} x (1-x) P_{gg}(x) \left(\frac{{\rm ln}(1-x)}{1-x}\right)_+ \, .
\eea
In \SCETbb, the partonic cross section takes on the factorized form of Eq.~(\ref{csmom}), except now the
initial state is a gluon rather than a nucleon:
\bea\label{cspart}
\frac{\hat\sigma^{II}_N}{\sigma_0} = x \,\tilde{S}^{(8,{}^1S_0)}(N) \int_x^1 \frac{d\xi}{\xi} \, \hat{C}_{II}\left(\frac{x}{\xi},N \right) f_{g/g}(\xi) \, .
\eea
In Eq.~(\ref{cspart}), we need the one loop result for the moments of the shape function,
\bea
S^{(8,^1S_0)}(N) = 1 - \frac{\alpha_s C_A}{\pi} \left[ \log^2 \left(\frac{\bar \mu \bar N}{M} \right) - \log \left(\frac{\bar \mu \bar N}{M} \right)
 \right] \, ,
\eea
and the one loop expression for the gluon structure function~\cite{Sterman:1994ce}
\bea
f_{g/g}(\xi) = \delta(1-\xi) -\frac{1}{\eps}\frac{\alpha_s}{2 \pi} {\cal P}_{gg}(\xi) \, .
\eea
Parametrizing the matching coefficient $\hat C_{II}$ as
\bea
\frac{x}{\xi}
\hat C_{II}\left(\frac{x}{\xi},N\right) = \delta\left(1-\frac{x}{\xi}\right)(1+ \alpha_s \hat C_{II}^{(1)}(N)) 
+\alpha_s \hat C^{(2)}_{II}\left(\frac{x}{\xi},N\right)\, ,
\eea
inserting into Eq.~(\ref{cspart}) and expanding to $O(\alpha_s)$ we find
\bea\label{scetIImom}
\frac{\hat\sigma^{II}_N}{\sigma_0} &=&\delta(1-x)\left[1+ \alpha_s \hat C_{II}^{(1)}(N) 
- \frac{\alpha_s C_A}{\pi} \left[ \log^2 \left(\frac{\bar \mu \bar N}{M} \right) - \log \left(\frac{\bar \mu \bar N}{M} \right)\right]\right]  \nn \\
&&  -\frac{1}{\eps}\frac{\alpha_s}{2 \pi} x {\cal P}_{gg}(x) + \alpha_s \hat C^{(2)}_{II}(x,N) \,.
\eea
Comparing Eq.~(\ref{scetImom}) and Eq.~(\ref{scetIImom}) we obtain the matching coefficient $C_{II}$ to $O(\alpha_s)$:
\bea
\hat C_{II}(x,N) &=& \delta(1-x) \left( 1 + \frac{C_A \alpha_s}{\pi}\left[\frac{1}{2}{\rm ln}^2\left(\frac{\bar \mu^2 
\bar N}{M^2}\right) -\frac{\pi^2}{8}\right]\right) \nn \\
&&- \frac{\alpha_s}{2\pi} {\overline P}_{gg}(x) 
{\rm ln}\left(\frac{\bar \mu^2 x \bar N}{M^2}\right)
+  \frac{\alpha_s}{2\pi} (1-x) P_{gg}(x) \left(\frac{{\rm ln}(1-x)}{1-x}\right)_+ \, . 
\eea
This result shows that all large logs of $N$ in the matching coefficient vanish at the intermediate scale $\bar \mu = M/\sqrt{\bar N}$, 
which fixes the \SCETa-\SCETb matching scale. At this scale there remains a ln$(x)$ in the matching coefficient which is $O(1)$ 
in our power counting.

\section{Running}

In this section the renormalization group is used to resum large logs of $N$ in the
moments of the cross section. An inverse Mellin transform of the resummed moments is
performed to obtain an analytic resummed formula for the differential cross section.
The evolution is carried out in two stages. First, we run the
effective theory currents in Eq.~(\ref{currmatch}) using the SCET${}_{\textrm{I}}$ 
RGEs from the scale $\bar \mu =M$, where QCD is matched onto SCET${}_{\textrm{I}}$, to the scale $\bar \mu =
M/\sqrt{\bar N}$, where SCET${}_{\textrm{I}}$ is matched onto SCET${}_{\textrm{II}}$. Second, the 
shape function in SCET${}_{\textrm{II}}$ is run down to the scale $\bar \mu=M/\bar N$. 

The currents that arise in photoproduction also arise in radiative decays, and the running was calculated in
Ref.~\cite{Bauer:2001rh}: 
\begin{eqnarray} \label{LOC}
 \log\bigg[\frac{C_V(\mu)}{C_V(M)}\bigg] &=& 
-\frac{4\pi\Gamma^{\rm adj}_1}{\beta_0^2\:\alpha_s(M)}
 \:\Big[ \frac{1}{y} -1 + \log y \Big]
 -\frac{\Gamma^{\rm adj}_1\beta_1}{\beta_0^3} 
\Big[ 1 -y + y\log y 
         -\frac12 \log^2 y \Big] 
\nonumber\\ 
 && - \frac{B_1+2 \gamma_1}{\beta_0} \log y
  - \frac{4 \Gamma^{\rm adj}_2}{\beta_0^2}  
  \Big[ y -1- \log y \Big] \,, 
\end{eqnarray}
where $y = \alpha_s(\mu)/\alpha_s(M)$ and
\begin{eqnarray}
\label{param}
\Gamma^{\rm adj}_1 &=&  C_A ,\quad 
\Gamma^{\rm adj}_2 =  
   C_A \left[ C_A \left( \frac{67}{36} - \frac{\pi^2}{12} \right) 
    - \frac{5n_f}{18} \right] , \nonumber \\
B_1 &=& -C_A\,, \quad \gamma_1 = -\frac{\beta_0}{4} \,,
\quad \beta_0 = \frac{11}{3}C_A -\frac{2}{3} n_f  \,.
\end{eqnarray}
The two loop cusp anomalous dimension $\Gamma^{\rm adj}_2$ was first calculated in Refs.~\cite{Korchemsky:1987ts,Korchemsky:1993xv}.
At leading order $C_V(M) =1$. 

In the second stage, only the shape function is evolved. This is possible since
the soft and collinear sectors in SCET${}_{\textrm{II}}$ are decoupled. Denoting the scales in the collinear and soft sectors as $\bar \mu_c$ and $\bar \mu_s$ respectively, and making the scales 
explicit in Eq.~(\ref{csc}), we have
\bea\label{scales}
\frac{d \sigma}{d z } &=&  \rho \, \sigma_0  C^2_V(\bar \mu_c)
\int d k^+  \, S^{(8,^1S_0)}(\sqrt{s}(1-z)-k^+;\bar \mu_s) \nn \\
& & \times
\int_\rho^1 \frac{d\xi}{\xi} \, C_{II}\Big(\frac{\rho}{\xi},k^+;\bar \mu_c\Big) 
f_{g/P}(\xi; \bar \mu_c ) \,.
\eea
In this section, we focus on the $^1S_0^{(8)}$ contribution to the cross section. The 
expression for the resummed cross section  is easily generalized to $^3P_J^{(8)}$  channels
since the evolution equations are identical. In matching \SCETa onto \SCETb we
must set $\bar \mu_s = \bar \mu_c = M/\sqrt{\bar{N}}$ in order to minimize logarithms in the
matching coefficient $C_{II}$. However, the shape function will still contain large logarithms
of $N$. The shape function must be evolved from $\bar \mu_s = M/\sqrt{\bar N}$ to $\bar \mu_s =M/\bar{N}$ to minimize
its logarithms. 

The running of the soft function is easily carried out in moment space where the differential cross section is a product of the 
moments of the shape function and moments of the hard coefficient, as in Eq.~(\ref{csmom}),
\begin{equation}\label{mom}
\sigma_N = \rho \, \sigma_0 C^2_V(\bar \mu_c)
\tilde{S}^{(8,^1S_0)}(N;\bar \mu_s) \int^1_\rho \frac{d\xi}{\xi}  \, 
\hat{C}_{II}\Big(\frac{\rho}{\xi} ,N;\bar \mu_c\Big) f_{g/P}(\xi; \bar \mu_c ) \,.
\end{equation}
The anomalous dimension of the shape function can be calculated by replacing the $\jpsi$ projection operator 
${\cal P}_\psi = a^\dagger_\psi a_\psi$ with an on-shell charm projection operator 
${\cal P}_{c\bar{c}}$ since the renormalization of an operator is only 
sensitive to short distances. The tree and one gluon Feynman rules are the same Feynman rules as those of the shape function 
that appears in quarkonium decay, and the one loop calculation of the shape function in Ref.~\cite{Bauer:2001rh} gives
\begin{equation}\label{softanomdim}
\gamma(N;\bar \mu) = \frac{\alpha_s(\bar \mu) C_A}{\pi} 
\Bigg[ 1 - 2 \log \Bigg( \frac{\bar \mu \bar{N}}{M} \Bigg) \Bigg] \,.
\end{equation}
The RGE can be solved in moment space. Combining the  running in SCET$_{\rm I}$ of the currents in 
Eq.~(\ref{LOC}) with the evolution of the shape function, we obtain  the following resummed expression for the moments of 
the $\jpsi$ photoproduction cross section:
\bea\label{fullyresummedN}
\sigma_N &=& \rho \, \sigma_0  \,  e^{\log(N) g_1(\chi) + g_2(\chi)} \tilde{S}^{(8,^1S_0)}(N;M/\bar{N}) 
\nn \\
&& \times
\int^1_\rho \frac{d\xi}{\xi} \, \hat{C}_{II}\Big(\frac{\rho}{\xi},N;M/\sqrt{\bar{N}}\Big) f_{g/P}(\xi;M/\sqrt{\bar{N}} ) \,,
\eea
where 
\begin{eqnarray}
\label{gis}
g_1(\chi) &=& 
-\frac{2 \Gamma^{\rm adj}_1}{\beta_0\chi}\left[(1-2\chi)\log(1-2\chi)
-2(1-\chi)\log(1-\chi)\right], \nonumber \\
g_2(\chi) &=& -\frac{8 \Gamma^{\rm adj}_2}{\beta_0^2}
  \left[-\log(1-2\chi)+2\log(1-\chi)\right] \nonumber\\
 && - \frac{2\Gamma^{\rm adj}_1\beta_1}{\beta_0^3}
   \left[\log(1-2\chi)-2\log(1-\chi)
  +\frac12\log^2(1-2\chi)-\log^2(1-\chi)\right] \nonumber\\
 &&+\frac{4\gamma_1}{\beta_0} \log(1-\chi) + 
 \frac{2B_1}{\beta_0} \log(1-2\chi)\nonumber\\
 & &  -\frac{4\Gamma^{\rm adj}_1}{\beta_0}\log n_0
 \left[\log(1-2\chi)-\log(1-\chi)\right]\,,
\end{eqnarray}
with $n_0 = e^{\gamma_E}$, $\chi=\log (N)\, \alpha_s(M)\beta_0/4\pi$, and 
$\beta_1 = (34C_A^2-10C_A n_f-6C_F n_f)/3$.

To obtain the resummed  expression for the differential cross section, $d\sigma/dz$, we take the inverse Mellin transform of the 
expression in Eq.~(\ref{fullyresummedN}).  However, this is complicated since the scale in the pdf depends on $N$. To perform
the inverse Mellin transform we must first undo the convolution in Eq.~(\ref{fullyresummedN}) by taking 
moments with respect to $\rho$:
\bea\label{doublemellin}
\frac{\tilde\sigma_N(K)}{\sigma_0} &=& \frac{1}{\sigma_0}  \int_0^1 d\rho\, \rho^{K-1} \sigma_N(\rho) \\
&=& e^{\log(N) g_1(\chi) + g_2(\chi)} \tilde{S}^{(8,{}^1S_0)}(N; M/\bar{N} )
\nn \\
& & \times 
\int_0^1 d\rho\, \rho^K \int_\rho^1 \frac{d\xi}{\xi} \, \hat{C}_{II}\left(\frac{\rho}{\xi},N;M/\sqrt{\bar{N}} \right) 
f_{g/P}(\xi;M/\sqrt{\bar{N}})\nn\\
&=& e^{\log(N) g_1(\chi) + g_2(\chi)} \, \tilde{S}^{(8,{}^1S_0)}(N; M/\bar{N}) \,
\tilde{C}_{II}\left(K,N;M/\sqrt{\bar{N}}  \right) \, \tilde{f}_{g/P}(K;M/\sqrt{\bar{N}} ), \nn
\eea
where
\bea
\tilde{C}_{II}\left(K,N; \mu\right) &=& \int_0^1 d\xi \,\xi^K \hat{C}_{II}\left(\xi,N ; \mu \right),\nn\\
\tilde{f}_{g/P}(K; \mu) &=& \int_0^1 d\xi \,\xi^K f_{g/P}(\xi ; \mu).
\eea
The dependence of the pdf on the scale $M/\sqrt{\bar{N}}$ can be made explicit by using the  evolution 
equations for the moments of the structure function. Ignoring mixing, the running of the pdf in moment space is given by  
\bea\label{DGLAP}
\mu\frac{d}{d\mu} \tilde{f}_{g/P}(K;\mu) = \frac{\alpha_s(\mu)}{4\pi} \, a_{gg}(K) \tilde{f}_{g/P}(K;\mu)\,,
\eea
where the explicit form of $a_{gg}(K)$ is not needed in what follows.  The leading order solution of Eq.~(\ref{DGLAP})
in moment space is
\bea\label{pdfrun}
\tilde{f}_{g/P}\left(K;M/\sqrt{\bar{N}}\right) = \tilde{f}_{g/P}(K;M)\exp\left[\frac{a_{gg}(K)}{2\beta_0} \log(1-\chi)\right] 
\equiv \tilde{f}_{g/P}(K;M)\exp\left[h_K(\chi)\right].
\eea
The pdf on the right hand side of this equation is evaluated at the scale $M$, 
and is therefore independent of $N$. All $N$ dependence has been moved into the factor $h_K(\chi)$. Using 
this result in Eq.(\ref{doublemellin}) we get
\bea\label{newresum2}
\frac{\tilde\sigma_N(K)}{\sigma_0} =  e^{\log(N) g_1(\chi) + g_2(\chi) + h_K(\chi)}
\tilde S^{(8,^1S_0)}(N;M/\bar{N}) \, \tilde C_{II}(K,N;M/\sqrt{\bar{N}}) \,\tilde{f}_{g/P}\left(K; M \right).
\eea
Since the logarithms in $\hat C_{II}(x,N)$ are minimized at the scale $M/\sqrt{\bar N}$,
$\tilde C_{II}(K,N;M/\sqrt{\bar{N}})= 1+O(\alpha_s)$, where the $O(\alpha_s)$ term has no logarithmically
enhanced contributions. Therefore, we can set $\tilde C_{II}(K,N;M/\sqrt{\bar{N}})= 1$, making
it possible to analytically evaluate the inverse Mellin transforms with respect to 
both $N$ and $K$. Using the results of Ref.~\cite{Leibovich:1999xf} to evaluate the inverse Mellin transform with respect $N$ yields
\bea\label{almost}
\frac{d \tilde\sigma(z, K)}{dz} &=& \sigma_0 \int_z^1 \frac{du}{u}\hat S^{(8,^1S_0)}(z/u)\nn\\
&&\times\left(-u\frac{d}{du}\left\{\Theta(1-u)\frac{e^{l g_1(l) + g_2(l) + h_K(l)}}{\Gamma[1-g_1(l) - l g_1'(l)]}\right\}\right)
\tilde{f}_{g/P}(K; M),
\eea
where $l = -\alpha_s\beta_0/(4\pi) \log(1-u)$ and $g_1'(l) = dg_1(l)/d l$. Next, we eliminate 
the factor $h_K(l)$ using the leading order solution of Eq.~(\ref{DGLAP}) again,
\bea
e^{h_K(l)} \tilde{f}_{g/P}(K; M) = \tilde{f}_{g/P}(K; M)\exp\left[\frac{a_{gg}(K)}{2\beta_0} \log(1-l)\right] = 
\tilde{f}_{g/P}\left(K; M\sqrt{1-u}\right).
\eea
Using this result in Eq.~(\ref{almost}) we arrive at
\bea
\frac{d \tilde\sigma(z, K)}{dz} &=& \sigma_0\int_z^1 \frac{du}{u}\hat S^{(8,^1S_0)}(z/u)\\
&&\times\left(-u\frac{d}{du}\left\{\Theta(1-u)\frac{e^{l g_1(l) + g_2(l) }}{\Gamma[1-g_1(l) - l g_1'(l)]}\,
\tilde{f}_{g/P}(K; M\sqrt{1-u})\right\}\right).\nn
\eea
The $K$ dependence is now entirely contained in the moments of the pdf, so the inverse Mellin transform with respect to $K$ is 
trivial. The fully resummed differential cross section is
\bea\label{finalresult}
\frac{d\sigma}{d z} &=&\rho \, \sigma_0 \int_z^1 \frac{du}{u}\hat S^{(8,^1S_0)}(z/u) \\
&& \times
\, \left(-u\frac{d}{du}\left\{\Theta(1-u)\frac{e^{l g_1(l) + g_2(l) }}{\Gamma[1-g_1(l) - l g_1'(l)]} \, f_{g/P}(\rho ; M\sqrt{1-u})\right\}\right).
\nn
\eea

This result is unusual because the scale of the pdf varies between $M\sqrt{1-z} \sim \sqrt{M\Lambda_{QCD}}$ and $0$ in
the convolution integral. It is not clear how to evaluate the pdf when the scale is below $\LQCD$. Another problem
arises in the resummed coefficient since the functions $g_i(l)$ blow up for $u$ sufficiently close to 1. If the
resummed exponent is expressed as an integral  over a running coupling this divergence is due to  the Landau pole. 
This problem commonly arises in resummed calculations and a prescription for dealing with the Landau pole is required
for phenomenological work. In what follows we cut off the integral at an upper limit, $u_{max} =0.93$, which is set by
the location of the Landau pole.  For this value of $u_{max}$ the lowest scale at which the pdf needs to be evaluated 
is 800 MeV, so the prescription for dealing with the Landau pole in the resummed exponent also fixes the problem
associated with the scale of the pdf. The difference between integrating to $u_{max}$ as opposed to one is formally 
the same order as power suppressed corrections and  therefore can be systematically
neglected~\cite{Leibovich:2001ra}.  Likewise, the difference between the prescription used in this paper and other
possible prescriptions is of order power suppressed corrections. An alternative approach which explicitly
avoids the Landau pole is given in Ref.~\cite{Bosch:2003fc}. 

\section{Phenomenology}

Before we can investigate the phenomenological consequences of our resummed cross section, we must determine the shape
functions.  Unfortunately not much is known about these  nonperturbative functions, although they also arise in $e^+ +
e^-\to J/\psi + X$ and electroproduction~\cite{Beneke:1997qw}. A fit of the shape  function to Belle~\cite{Abe:2001za}
and Babar~\cite{Aubert:2001pd} data on $e^+ + e^- \to J/\psi + X$  was carried out in Ref.~\cite{Fleming:2003gt}, and
we will use the parameters determined from that fit in our  calculation.  We use these shape functions for
illustrative purposes only, as the shape functions extracted in Ref.~\cite{Fleming:2003gt} are not reliable.  The
Belle collaboration observes that  the $J/\psi$ cross section at $\sqrt{s} =10.6$ GeV is dominated by events with an
additional  $c \bar c$ pair, i.e., $e^+ + e^- \to J/\psi + c+ \bar c +X$~\cite{Abe:2002rb, Abe:2004ww}. This fact is
poorly understood at the present time \cite{Cho:1996cg} and we do not expect the  color-octet mechanism studied in
this paper to contribute significantly to $e^+ + e^- \to J/\psi + c + \bar c +X$. A sensible extraction of  the
color-octet shape function from $e^+ + e^- \to J/\psi + X$ requires removal of events with extra $c\bar c$ pairs, and
such data is not currently available.

For simplicity, we assume that the $^1S_0^{(8)}$ and $^3P_J^{(8)}$ shape functions are the same, but this need not be the case. The shape function model we adopt is a modified version of a model used in the decay of $B$
mesons~\cite{Leibovich:2002ys},
\begin{eqnarray}\label{sfmodel}
f(r^+) = \frac{1}{\bar\Lambda} \frac{a^{ab}}{\Gamma (ab)} (x -1)^{ab - 1} 
e^{-a(x-1)} \, \theta(x-1)  \,, \hspace{10ex} x = \frac{r^+}{ \bar\Lambda} \,,
\end{eqnarray}
where $a$ and $b$ are adjustable parameters and $\bar\Lambda= M_\psi - M$. This function is related to the shape function  {\it only} in the $J/\psi$ rest frame where $f(-r^+) =S(r^+)$. To relate the model above to the shape function $\hat{S}(u)$ which appears in Eq.~(\ref{finalresult}) we use boost invariance 
\begin{equation}
\label{boostsf}
dk^+ S(-\sqrt{s}(1-z) + k^+) \to dl^+S(-M(1-z) + l^+) \,,
\end{equation}
where $l^+=(M/\sqrt{s})k^+ $ is the rest frame residual momentum which is $O(\LQCD)$. Since $dk^+ =  (\sqrt{s}/M)d l^+ $ we get $S(-\sqrt{s}(1-z) + k^+) \to (M/\sqrt{s}) S(-M(1-z) + l^+)$.  Finally we get
\bea
\hat{S}(z-u) &=& \sqrt{s} \, S(-\sqrt{s}(1-z) + k^+)
\nn \\
&\to& M S(-M(1-z) + l^+)
\nn \\
&=& M f(M(1-z) -l^+) = M f(M(u-z)) \,,
\eea
where in the $\jpsi$ rest frame $l^+ \equiv M(1-u)$.

The first three moments  $ f(r^+)$ are
\begin{eqnarray}
m_0 = \int_{\bar\Lambda}^\infty dr^+\, f(r^+) =  1, & &  
m_1 = \int_{\bar\Lambda}^\infty dr^+\, r^+\, f(r^+) =   \bar\Lambda (b+1), \nn \\
m_2 = \int_{\bar\Lambda}^\infty dr^+\, (r^+)^2\, f(r^+) &=& 
   \bar\Lambda^2 \bigg( \frac{b}{a} + (b+1)^2 \bigg) \,.
\end{eqnarray}
Since $\bar\Lambda \sim {\cal O}(\Lambda_{\rm QCD} )$, any choice with  $a \sim b \sim {\cal O}(1)$ gives the desired scaling for the
moments.  We use parameters taken from a fit to the  $e^+ + e^-\to J/\psi + X$ data \cite{Fleming:2003gt}: $a=1, b=2$.
The value of the first and second moments of the shape function for this choice
of parameters are $890 {\rm\ MeV}$ and $(985 {\rm\ MeV})^2$ respectively.  Since $m_c v^2 \approx 500$ MeV the moments are consistent
with the velocity scaling rules.

\begin{figure}
\begin{center}
\includegraphics[width=6.25in]{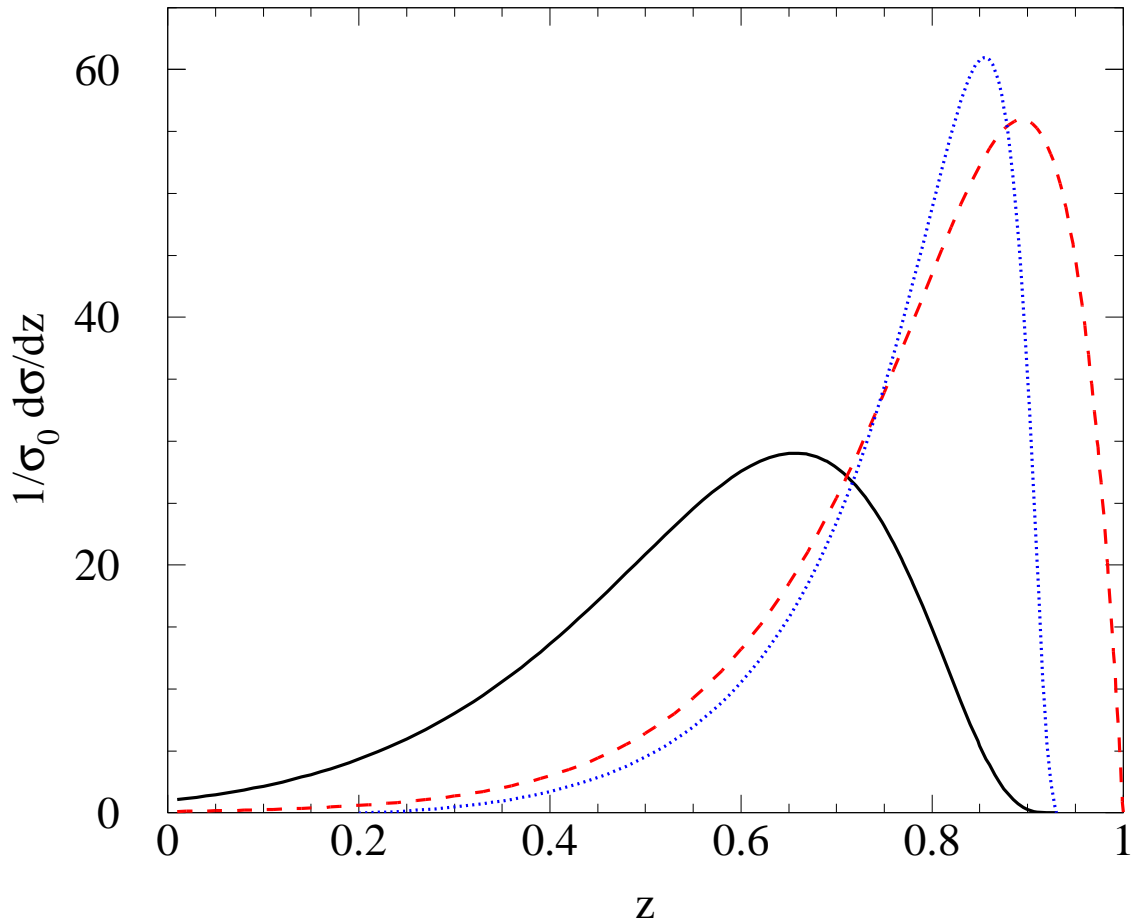}
\end{center}
\caption{\it  The color-octet contribution at the endpoint normalized to $\sigma_0$. The solid line is the perturbative resummation 
convoluted with the shape function. 
The dotted line is perturbative resummation only, and the dashed line is no resummation, but including a shape function.} 
\label{plotoctet}
\end{figure}

Calculations of $d\sigma/dz$ are shown in Fig.~\ref{plotoctet}. The solid line is the final result,
Eq.~(\ref{finalresult}), normalized to $\sigma_0$ using the shape function in  Eq.~(\ref{sfmodel}).
The dashed line is a plot of the shape function alone and the dotted line includes only the perturbative resummation.
Here we use $m_c = 1.4$ GeV, $\sqrt{s} = 100$ GeV, and the CTEQ5L pdfs~\cite{Lai:1999wy}.   Fig.~\ref{plotoctet} shows the effects of
both the perturbative resummation and the shape function.  The results are very similar to other resummed
calculations: the perturbative resummation alone or the shape function alone give a cross section that is highly
peaked in the endpoint region. In contrast,  the convoluted result is much broader with the peak shifted to
lower values of $z$. 

It is important to point out that the calculation is not valid for $z$ very close to one. Though SCET is valid in the
endpoint region, the calculation assumes that the final state is inclusive. However, as $z$ gets very close to one the
cross section is dominated by exclusive final states. This occurs when $M(1-z) \sim \LQCD$, which is roughly $z \sim
0.9$. In this exclusive region our analysis is not valid and a very different approach is needed. 

\begin{figure}
\begin{center}
\includegraphics[width=6.25in]{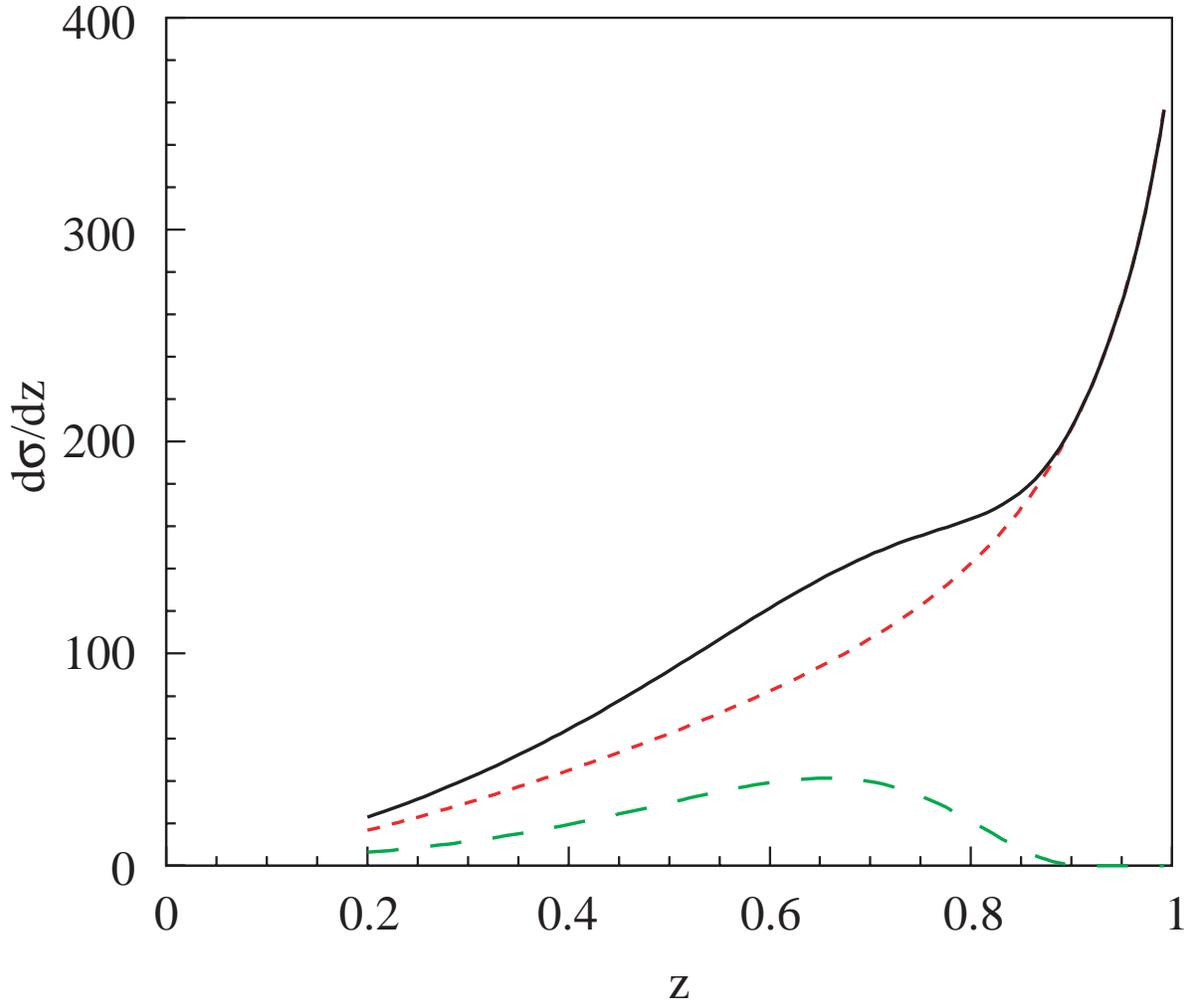}
\end{center}
\caption{\it  The differential cross section in nb. The resummed color-octet contribution (short dashes), the leading order color-singlet contribution (long dashes), and the sum (solid line). The rise in the differential cross section at the endpoint due to the color-singlet contribution must be resummed before a meaningful comparison to data can be made.} 
\label{plotsinglet}
\end{figure}

In Fig.~\ref{plotsinglet} we show the differential cross section. The short dashed line is the
color-octet contribution, which has a normalization set by the linear combination of color-octet matrix elements in
Eq.~(\ref{partcs}). We set this combination to $3 \cdot 10^{-3}$, which is an order of magnitude smaller than the
value determined from a fit to Tevatron data.  The long dashed curve is the color-singlet contribution, and the solid
curve is the sum of singlet and octet. The singlet contribution is peaked in the endpoint and it is not possible to
compare to data until a resummation of the singlet contribution is completed. This work is in progress~\cite{FLM}.

The analysis of Ref.~\cite{Cacciari:1996dg} indicates that the data is well described by the  color-singlet cross
section alone so it appears that there is little room for a color-octet contribution. A  naive  fit with the
leading-order color-singlet calculation and the resummed color-octet cross  section would lead to the conclusion that
color-octet matrix elements are smaller than those  obtained from fitting to Tevatron data. However there are caveats.
First, large endpoint effects in color-singlet production also need to be resummed. While we do not expect the results
of this resummation to be as dramatic as in color-octet production, resummation should also suppress color-singlet 
production in the endpoint region,  creating more room for a color-octet contribution to photoproduction. Second, most
of  the available data  has cuts that remove any contribution for $p_\perp < 1$ GeV.  In our calculation we integrated
over $p_\perp$ assuming $p_\perp < M$.  Thus care must be taken when comparing to experimental measurements. The Zeus
collaboration has carried out an analysis which includes a measurement of $d\sigma/dz$ with no $p_\perp$ cut as well
as a cut of 1~GeV and 2~GeV~\cite{Chekanov:2002at} . The data  clearly show that a $p_\perp$ cut decreases the
differential cross section with the greatest reduction being in the high $z$ bins. Thus comparing to data with no
$p_\perp$ cut may allow for a larger color-octet contribution.
Furthermore it is common practice to cut out the
diffractive contribution from the data.  However, our calculation assumes that the diffractive contribution is part of
the complete sum over final states, and an honest comparison requires data without the diffractive contribution
subtracted. The Zeus analysis of Ref.~\cite{Chekanov:2002at} gives the percentage by which the cross section decreases
when the diffractive contribution is removed. Again  this fraction becomes large as the endpoint region is approached,
and including the diffractive  contribution in the experimental measurement could lead to greater room for a
color-octet contribution.

\section{Summary}

In this paper we studied the color-octet contribution to $\jpsi$ photoproduction near the kinematic  endpoint. As $z
\to 1$ the usual NRQCD factorization formalism breaks down due to large perturbative and nonperturbative corrections.
We combined SCET with NRQCD to derive a factorizatioin theorem for the differential cross section, $d\sigma/dz$, in
the endpoint region in terms of the parton distribution function and nonperturbative color-octet shape functions.
Large Sudakov logarithms which appear in the endpoint are resummed using the RGE of the effective theory. 

Since the total photoproduction cross section is largely accounted for by the color-singlet term in NRQCD a
 resummation of this contribution must be carried out before a quantitative comparison to data can be made.  However,
 some qualitative conclusions can be drawn. First, we find that perturbative resummation acts constructively with the
 shape function to significantly broaden the $z$ distribution. This important effect improves the agreement between
 the shape of the theoretical prediction and data. Second, the normalization of our prediction is set by a linear
 combination of NRQCD matrix elements which also appear in calculations of $\jpsi$ production at the Tevatron. The
 data clearly prefer values for the color-octet matrix elements that are roughly an order of magnitude smaller than
 the central values extracted from Tevatron data. However, serious comparison with data requires resummation of large
 endpoint corrections to the color-singlet production cross section. Also, the assumptions underlying our calculations
 require that we apply our results to  data without cuts on $p_\perp$ and diffractive contributions that are commonly
 used in existing experimental analyses. Resumming color-singlet contribution and relaxing experimental cuts could
 both lead to more room for color-octet contributions in a complete analysis of $J/\psi$ photoproduction
 near $z \to 1$.

\acknowledgments 
A.L.~was supported in part by the National Science 
Foundation under Grants No.~PHY-0244599 and PHY-0546143.  Adam Leibovich is a Cottrell Scholar of the Research Corporation.  T.M.~was supported in part by the Department
of Energy under grant numbers DE-FG02-96ER40945 and DE-AC05-84ER40150.
\appendix 

\section{}

Here we derive Eqs.~(\ref{tr1}) and (\ref{tr2}). We aim to extract terms that are 
singular as $z \to 1$, so we  will systematically drop all contributions 
which are regular in this limit. We start with Eq.~(\ref{tr1}):
\bea\label{int1}
(1-z)^{-1-\eps}\int_0^z dx \frac{(z-x)^{-\eps}}{1-x} g(x) &=& \\ 
&&\hspace{-1.0 in}(1-z)^{-1-\eps}\left[g(1) \int_0^z dx \frac{(z-x)^{-\eps}}{1-x}   
+ \int_0^z dx \frac{(z-x)^{-\eps}}{1-x} (g(x)-g(1)) \right] \,. \nn
\eea
We assume $g(x)$ can be expanded about $x=1$. To evaluate the second term  on the right hand side of
Eq.~(\ref{int1}), we note the distributional identity in Eq.~(\ref{stan}) implies 
\bea 
(1-z)^{-1-\eps} f(z) = (1-z)^{-1-\eps} f(1) + ...\,,
\eea
where the ellipsis denotes terms that are not nonsingular as $z \to 1$ 
for any function $f(z)$ which can be expanded about  $z = 1$. The integral in the 
second term can therefore be evaluated with $z = 1$:
\bea
(1-z)^{-1-\eps} \int_0^z dx \frac{(z-x)^{-\eps}}{1-x} &&\hspace{-0.25 in}(g(x)-g(1))  \\
&=&(1-z)^{-1-\eps} \int_0^1 dx  (1-x)^{-1-\eps}  (g(x)-g(1)) \nn \\
 &=& (1-z)^{-1-\eps} \int_0^1 dx \left[ \left(\frac{1}{1-x}\right)_+ 
- \eps \left(\frac{{\rm ln}(1-x)}{1-x}\right)_+ \right] g(x) .\nn 
\eea
We can use the distributional identity in Eq.~(\ref{stan}) and expand in $\eps$ to
obtain
\bea\label{int1b}
(1-z)^{-1-\eps} \int_0^z dx  \frac{(z-x)^{-\eps}}{1-x}&& \!\!  (g(x)-g(1)) =   \\ 
&&\hspace{-1.0 in}\delta(1-z)\left[  -\frac{1}{\eps}\int_0^1 dx \left(\frac{1}{1-x}\right)_+ g(x) 
+ \int_0^1 dx \left(\frac{{\rm ln}(1-x)}{1-x}\right)_+ g(x) \right] \nn \\
&& \hspace{-1.0 in}
+\left(\frac{1}{1-z}\right)_+ \int_0^1 dx \left(\frac{1}{1-x}\right)_+ g(x) \, . \nn 
\eea
To evaluate the first integral on the right hand side of Eq.~(\ref{int1}) 
we remember that 
\bea
(1-z)^{-1-\eps}\int^z_0 dx \frac{(z-x)^{-\eps}}{1-x} \nn
\eea
is really a distribution so we consider the double integral:
\bea
I = \int_0^1 dz f(z) (1-z)^{-1-\eps}\int^z_0 dx \frac{(z-x)^{-\eps}}{1-x} \nn \, ,
\eea
where $f(z)$ is a smooth function. Writing $I$ as
\bea\label{tffa}
I &=&\int_0^1 dz [f(z)-f(1)] (1-z)^{-1-\eps}\int^z_0 dx \frac{(z-x)^{-\eps}}{1-x}
\nn \\
&& + f(1)\int_0^1 dz (1-z)^{-1-\eps}\int^z_0 dx \frac{(z-x)^{-\eps}}{1-x} \, ,\nn
\eea
we see that the integrand of the first term is finite as $z \to 1$ so we can set $\eps \to 0$
in evaluating this term. The second term is straightforward to evaluate after interchanging
the order of the $x$ and $z$ integrations. The result is  
\bea
I &=& -\int_0^1 dz [f(z)-f(1)] \frac{{\rm ln}(1-z)}{1-z} 
- f(1)\frac{1}{2\eps}\frac{\Gamma[-\eps] \Gamma[1-\eps]}{\Gamma[1-2\eps]} \nn \\
&=& \int_0^1 dz f(z)\left[\delta(1-z)\left(\frac{1}{2\eps^2} -\frac{\pi^2}{12}\right)
-\left( \frac{{\rm ln}(1-z)}{1-z} \right)_+ \right] \, ,
\eea
so we obtain the distributional identity
\bea\label{int1a}
(1-z)^{-1-\eps}\int^z_0 dx \frac{(z-x)^{-\eps}}{1-x}
 = \delta(1-z)\left( \frac{1}{2\eps^2}  -\frac{\pi^2}{12}\right) 
-\left( \frac{{\rm ln}(1-z)}{1-z} \right)_+ \, .
\eea
Combining Eq.(\ref{int1}), Eq.(\ref{int1a}), and Eq.(\ref{int1b}) yields the result in Eq.~(\ref{tr1}).

To obtain the result in Eq.~(\ref{tr2}), we first note that we can replace $g(x)$ with $g(1)$
and then use
\bea
\int_0^1 dz \, f(z)&&\!\!  \hspace{-0.1 in} (1-z)^{-\eps}\int_0^z dx \frac{(z-x)^{-\eps}}{(1-x)^2}  \nn \\
&& = \int_0^1 dz [f(z)-f(1)]  \int_0^z dx \frac{1}{(1-x)^2} +
f(1) \int_0^1 dz (1-z)^{-\eps}  \int_0^z dx \frac{ (z-x)^{-\eps}}{(1-x)^2} \nn \\
&& = \int_0^1 dz f(z) \left[ -\frac{1}{2\eps}\delta(1-z) + \left(\frac{1}{1-z}\right)_+ \right] \, ,
\eea
plus nonsingular terms.

\section{}

Here we discuss zero-bin subtractions in more detail and show how they can be implemented at the level of the Lagrangian by introducing a 
fictitious field. 	
A collinear particle is defined in SCET by  rephasing the full theory fields: 
\bea\label{field}
\phi(x) = \sum_{\vec p \neq 0} e^{-i \vec{p}\cdot \vec{x}} \phi_{\vec p}(x) \, .
\eea
Loop integrals as well as phase space integrations involve both a sum over labels
and an integral over residual momentum which can be performed using the following trick:
\bea\label{inttrick}
\sum_{\vec{p}} \int d^D k = \int d^D p \, ,
\eea
which allows one to avoid doing explicit sums in loops and in phase space.
Note that the sums in Eq.~(\ref{field}) and Eq.~(\ref{inttrick}) differ. In defining the fields, 
we only include nonvanishing label momentum, $\vec{p} \neq 0$, since when the label momentum 
vanishes the particle should really be regarded as ultrasoft. However, the trick in Eq.~(\ref{inttrick})
requires that the sum include the so-called ``zero-bin'', the momentum region with vanishing 
label momentum $\vec{p} = 0$. Careful evaluation of the sum over labels and integrals
requires modifying Eq.~(\ref{inttrick})~\cite{Manohar:2006nz}:
\bea\label{sub}
\sum_{\vec{p}\neq 0} \int d^D k \, F[p,k] = \int d^D p \left(F[p]- {\rm subtraction} \right) \, .
\eea

In a tree level calculation which involves phase space integrals,
one can deal with the zero-bin  by putting a cutoff in the phase space integral that separates the integration region  into $x-z \sim
O(\lambda^2)$ and $x-z \sim O(1)$ regions, and only use the appropriate mode in the respective  regions. This is
somewhat clumsy and a more elegant way to acheive the desired result is to extend both the collinear and soft
particle diagrams to cover the entire phase space and introduce a fictitious particle, $\phi^{\rm zb}$, which cancels off the double
counting in the corner of phase where such a cancellation is needed.
We modify the definition of the collinear field so the sum is over all labels
but the contribution of the spurious zero-mode is cancelled by a fictitious degree 
of freedom:
\bea\label{zmfield}
\phi(x) = \sum_{\vec p} e^{-i \vec{p}\cdot \vec{x}} \phi_{\vec p}(x) - \phi^{\rm zb}(x) \, .
\eea
If we use the fictitious $\phi^{zb}$ field then we can convert sums over labels 
ordinary loop integrals and phase space intergrals using Eq.~(\ref{inttrick}) but 
there are extra diagrams with the zero-bin field which cancel spurious zero-bin contributions in the
collinear graphs. These fields have the same couplings as collinear fields but their diagrams
contribute with opposite sign and they have vanishing  label momentum. Evaluating real emission graphs
with $\phi^{zb}$ yields Eq.~(\ref{zb}).


\end{document}